# Learning Analytics in Higher Education – Exploring Students' and Teachers' Expectations in Germany

Birthe Fritz[1], Dana Kube[2], Sonja Scherer[3], Hendrik Drachsler[4]

Abstract: Technology enhanced learning analytics has the potential to play a significant role in higher education in the future. Opinions and expectations towards technology and learning analytics, thus, are vital to consider for institutional developments in higher education institutions. The SHEILA framework offers instruments to yield exploratory knowledge about stakeholder aspirations towards technology, such as learning analytics in higher education. The sample of the study consists of students (N = 1169) and teachers (N = 497) at a higher education institution in Germany. Using self-report questionnaires, we assessed students' and teachers' attitudes towards learning analytics in higher education teaching, comparing ideal and expected circumstances. We report results on the attitudes of students, teachers, as well as comparisons of the two groups and different disciplines. We discuss the results with regard to practical implications for the implementation and further developments of learning analytics in higher education.

## 1. INTRO

Advanced digital education interventions, such as learning analytics (LA) can support making sense of learning and teaching data for better educational decision-making. The measurement, collection, analysis and reporting of big data about learners and teachers in the classroom can facilitate understanding and optimize the teaching and learning processes with real-time individualized feedback and support [23]. The implementation of supportive technologies, though is different in a country as Germany, compared to Australia or the USA, for instance, where it is more common to be used in educational settings. Studies in German-speaking higher education institutions (HEI) have revealed that only a very limited number of academic staff is using digital technologies in their courses [32, 3, 25]. An investment in technological equipment for the universities seems insufficient to address these gaps, though, as personal barriers such as deficits in e-competence have been shown to be significant factors for the adoption of technologies [29]. In the implementation process of advanced technological service based on methods of artificial intelligence (AI), machine learning or analyses of algorithms, there are barriers related to ethics and privacy concerns. For example, the so-called privacy paradox, which describes an apparent logical gap between peoples' desires and what they are willing to do regarding privacy may be an issue [38, 16]. Contradicting sentiments might not always lead to contradicting behaviour; however, it shows the relevance of considering the organization's socio-cultural context when estimating terms of change. To address issues with the implementation of LA, the SHEILA project (Supporting Higher Education to Integrate Learning Analytics) was initiated [36]. The SHEILA project offers a framework to guide HEI with the implementation and adoption of LA policies and strategies [35]. The work of Wollny et al. (2023) recently reported on applying early steps of the SHEILA framework, investigating the attitudes of a small sample of students towards digitalization at a German-speaking university in comparison to European SHEILA data samples from UK, Netherlands, and Estonia [42]. There were some tendencies among the students across Europe that LA is seen as ideal in HEI, however on average less realistic to be implemented that way. This study has both theoretical and practical aims: The empirical investigation of stakeholders' opinions on LA are supposed to serve the creation of a stakeholder engagement strategy for LA, especially AI-based LA, at the institution. First, we report on stakeholders' opinions on LA at Goethe University Frankfurt. Secondly, we dig deeper into empirical findings, comparatively identifying differences and similarities among students and teachers. The following research questions will thus guide our study:

---

[1] Goethe University, Institute of Psychology, Theodor-W.-Adorno-Platz 6, 60629, Frankfurt am Main, Germany & Johannes Gutenberg-University Mainz, Department of Clinical Psychology and Neuropsychology, Wallstraße 3, 55122 Mainz, Germany, birthe.fritz@stud.uni-frankfurt.de, https://orcid.org/0009-0005-4967-9186

[2] Goethe University, studiumdigitale, Eschersheimer Landstraße 155-157, 60323, Frankfurt am Main, Germany, kube@studiumdigitale.uni-frankfurt.de, https://orcid.org/0000-0002-8969-1869

[3] Goethe University, Institute of Psychology, Theodor-W.-Adorno-Platz 6, 60629, Frankfurt am Main, Germany, scherer@psych.uni-frankfurt.de, https://orcid.org/0000-0001-7063-2976

[4] Goethe University, studiumdigitale, Eschersheimer Landstraße 155-157, 60323, Frankfurt am Main, Germany, drachsler@studiumdigitale.uni-frankfurt.de, https://orcid.org/0000-0001-8407-5314

### 1.1 Research Questions

RQ1. What are the expectations of university students towards LA?

RQ2. What are the expectations of university teachers towards LA?

RQ3. Are there differences between the expectations of students and teachers towards LA?

#### 1.2 Learning Analytics and Artificial Intelligence

Among various trending topics that have gained ample attention recently, both in the media and academia, there is a highly pronounced demand of making use of educational data with AI to enhance the entire educational ecosystem. Within the realm of educational technology research, Learning Analytics (LA) serves as the overarching framework for investigations into the integration of AI within learning processes. LA may include methods of AI and deals with the assessment, collection and analysis of learners' data, with the objection to support reflection and provide prediction on learning outcomes [11]. LA does not only provide overviews on learning progress and achievements in form of summative feedback, but it also allows adaptive, personalised learning experiences by integrating formative assessments and feedback [6]. Researchers and practitioners concerned with LA have pointed out that the success of LA does not only depend on the characteristics of the application, but in the end on whether faculty and staff make use of it [24, 13]. Major challenges in the implementation of LA that have been identified in previous research address three main categories: technical, educational, and social challenges [42]. Whereas *technical* challenges refer to the scalability of services or data lake infrastructures, and *educational* challenges refer to reliability [31] and applicability [1], the third category of *social* challenges includes the integration of the interests an institution's stakeholders [10, 42]. Taking the *social infrastructure* and all relevant *stakeholders* of an institution into account has also been recommended by other researchers concerned with successful implementations of LA [19]. In this context, research indicated that different stakeholders differ in their expectations towards LA [14, 12]. Furthermore, it has been shown that differentiating between *ideal expectations* and *predicted expectations* is helpful for investigating ambiguities in attitudes or beliefs [37, 40]. In this context, stakeholders' judgments pertain to both *ideal*, desired outcomes and *predicted*, expected outcomes. The latter can be defined as the user's estimation of what is most likely to happen. [40]. The literature review by Mahmoud et al. (2020) confirms that stakeholders' views on LA vary [24]. So far, most research studies have focused on singular groups of stakeholders, though, such as students, teachers, or faculty. The importance of integrating the expectations of both students and teaching staff into the preparation of LA implementation has been emphasized as a result of quantitative and qualitative research [24].

Recent research has started to take different stakeholders' opinions into account [14, 12, 26, 7]. However, large-scale studies that compare different stakeholders' expectations are only in their infancy.

### 1.3 Stakeholders' Perspectives on Learning Analytics

Stakeholders involved in LA implementation or use at HEIs differ in their backgrounds, experiences, motives and strategic interests. So, it is not surprising that stakeholders also differ in their attitudes and priorities concerning LA. On the one hand, experts around the world have highlighted that the interests of all stakeholders should be taken into account in strategies for implementing LA successfully [13, 27]. On the other hand, finding common ground seems to be challenging. In research on LA, students are the stakeholders of most interest so far [24]. As the field of research is rather new [42], early studies have started with exploratory investigations [16], using mostly qualitative methods such as interviews [18, 33], focus groups [28] or content analyses [34] in order to identify main aspects that play a role for students to accept and adapt LA. More recent studies have used quantitative methods such as surveys and yielded greater sample sizes [38]. The most important student expectations on LA indicated by the studies mentioned here may be summarized by students' desires and fears: students reported desiring personalization, prompt feedback and a tracking record of their performance; they reported to be worried about data misuse and about the comparison of their academic performance with the ones of their fellow students [24]. To some extent, student needs were shown to be contradictory [34, 38]. Besides students, faculty is also an increasingly investigated stakeholder in LA processes [24]. Exploratory studies offer a broad array of ideas that faculty members have about LA, retrieved by interviews [17, 21, 30, 9], focus groups [21] or content analyses [34]. Quantitative studies on faculty's expectations are rarely up to date [22, 39]. One of the most striking results indicated by qualitative and quantitative studies alike is that teachers at HEI seem to appraise the potential of LA to support learning and teaching processes, but do not necessarily feel obliged to act based on data provided by LA. In general, faculty agrees that LA is no fast-selling item, as successful LA adoption can be affected by 'resources, technical skills, trustworthiness, data availability, accessibility, organisational culture, integration, and technological infrastructure' ([24], p.42). Research also highlights the importance of recognizing possible gaps between ideal and realistic expectations, showing that faculty report not being convinced that LA will be realized in an ideal way at their HEI [22]. To this date, there are only a few studies that integrate data from different stakeholders, going beyond an analysis of focus groups, interviews, and case studies [7, 20]. Hilliger et al. (2020) conducted a mixed methods study in four Latin American HEIs in which three groups of stakeholders were surveyed, showing that '(1) students need quality feedback and timely support; (2) teaching staff need timely alerts and meaningful performance evaluations, and (3) managers need quality information to implement support interventions' ([14], p.1). Stakeholders' differences are for instance disagreement on using student data for the improvement of learner agency [12]. For example, students did not comply with teaching staff aspiring an extended access to their data, reporting privacy concerns [12, 26]. In contrast, study results showed that students were optimistic about HEI teachers being able to interpret provided data, whereas teachers themselves reported worries about being unable to cope with the data and about increased workload [12]. In our study, we focus on the gap mentioned above between ideal and predicted expectations of both students and teaching staff, aiming at shedding light on possible weak points within the institution. Based on long-standing European experiences with implementing LA frameworks such as SHEILA [37], we aim to reach a shared vision on how to best implement LA at our HEI through the stakeholder engagement process.

## 2 METHODS

### 2.1 Procedure & Participants

As part of the Goethe University Frankfurt's central AI projects (ALI and IMPACT), a survey on the acceptance and attitudes of students and teachers to the use of LA in teaching was conducted via central email distribution lists in July-August 2022. The Data was collected within two months. Students ($n$ = 1169) came from all 16 departments. The results of those who answered the socio-demographic questions are shown in Table 1. The socio-demographic data of those teachers who answered these questions (*n* = 497), came from all 16 departments and are also shown in Table 1.

Table 1: Sociodemographic data of the students and teacher survey

| | | Students | | | Teachers | |
|---|---|---|---|---|---|---|
| | | % | *n* | | % | *n* |
| Level of education | Bachelor level | 42.2% | 493 | Scientific staff (post-doc) | 23.74% | 118 |
| | Master level | 19.9% | 233 | Lecturers | 20.52% | 102 |
| | Similar degree | 27.7% | 323 | Scientific staff (PhD candidate) | 17.71% | 88 |
| | | | | Academic councillors | 3.82% | 19 |
| | | | | Online-/ media-didactic staff | 1.61% | 8 |
| | PhD | 4.4% | 51 | Professors | 26.76% | 133 |
| Faculty | STEM | 31.9% | 373 | | 38.6 % | 192 |
| | Social science & humanities | 58.7% | 686 | | 50.3% | 250 |
| Gender | Women | 57.2% | 669 | | 38.8% | 193 |
| | Men | 32.6% | 381 | | 46.7% | 232 |
| | Divers | 2% | 23 | | 0.8% | 4 |
| Country | EU | 85% | 994 | | | |
| | Outside EU | 9% | 105 | | | |
| Working besides studying | Yes | 54.5% | 637 | | | |
| | No | 39.6% | 463 | | | |
| | *M (SD)* | *n* | | | *M (SD)* | *n* |
| Age | 25.98 years (SD = 8.9) | 1091 | | Teaching experience | 14.44 years (SD = 10.6) | 448 |

### 2.2 Instruments & Measures

**Student survey.** The student survey ('Student Expectations of Learning Analytics Questionnaire (SELAQ)') measures ideal expectations ('what students desire') and predicted expectations ('what students expect in reality') of LA. It is based on the SHEILA project, which developed a framework to promote formative assessment and personalized learning at HEIs through the direct engagement of stakeholders in the development process [36, 35]. The student survey (see all items in Table 6 in the annex) consists of twelve questions about LA service expectations (beliefs about the likelihood that future implementations and running of LA services will possess certain features). Each item assesses on two expectations 7-point Likert-scales, the (1) ideal and the (2) predicted value of questions regarding data security, consent, decision making, learning goals, feedback, skill development and the obligation to act. The variation of the items is explained by a two-factor structure of 'ethical and privacy expectations' (EP) and 'service feature expectations' (SF). EP pertains to student attitudes about the ethical practices involved in LA services. SF refers to how students would like to benefit from LA services [40]. The survey is a standard instrument for assessing student opinion in LA [41, 42]. It was validated using an iterative technique, in which factor analysis – exploratory and confirmatory – was used to refine the number of items and assess the validity of the scales. Other analyses, such as evaluating measurement invariance and latent class analysis, were also carried out [35].

**Teacher Survey.** Like SELAQ, the teacher survey includes two expectation Likert-scales, each measuring ideal expectations and predicted ones regarding LA at HEIs per item. It consists of an extended version of the original survey, amended from 16 to 20 questions, based on the same framework adopted to develop the student survey regarding SF and EP of LA, however extending the survey questions regarding specific teaching matters, so only seven items were identical and comparable between the two surveys (please see Tables 7 and 8 in the annex for details). Based on a former version of the survey by Kollom et al. (2021), it was redesigned and distributed among the Ruhr-University Bochum [22]. Given that a study has yet to validate the survey instrument in German and English, no reports about factor analytics have been published yet—only descriptive statistics. Our study will address this issue in the results section (see point 3).

### 2.3 Data Analysis

All quantitative descriptive explorative data analysis was conducted in SPSS® Statistics (Version 27, IBM® 2023). We conducted a PCA (Principal Components Analysis) under the preconditions for students and teachers that we had given sufficient sample size, linearity, freedom from outliers, and continuous variables. The Kaiser–Meyer–Olkin measure verified the sampling adequacy for the analysis. Only those factors were considered whose eigenvalues met the Kaiser's criterion (>1) with accompanying observation of the scree plot [8]. Continuous data is represented as mean (*M*) and standard deviation (*SD*), discrete data in frequencies and percentages. Statistical significance was established at $\alpha \leq .05$.

## 3 RESULTS

### 3.1 Factor Analysis Student Survey

In order to show the validity of the student survey instrument, we conducted a Principal Components Analysis (PCA) based on a correlation matrix to extract the most important independent factors with a value higher than one, with the values from the Ideal Case scale. The Kaiser–Meyer–Olkin measure of sampling adequacy was in the students' sample .87 and in the teacher's sample .95. The student survey results in the assessment of Kaiser's criteria and the scree-plot showed that the varimax-rotation verified the survey's two-factor (Ethics & Privacy, Service Features) solution with eigenvalues exceeding 1, explaining 60.87% of the total variance. Good results of the reliability analysis accompanied this ($\alpha = .83$), including both survey content domains, 'Ethics and Privacy' factor ($\alpha = .83$) and 'Service Features' factor ($\alpha = .89$). See annex Table 9 for details.

### 3.2 Factor Analysis Teacher Survey

To analyse the factor analysis of the staff survey, we also conducted a PCA based on a correlation matrix to extract the most important independent factors with a value higher than one, with the values from the ideal case scale. Results of Kaiser's criteria and scree-plot showed that the varimax-rotation verified the survey's three-factor (SF for teachers, SF for students and EP - teachers) solution with eigenvalues exceeding 1, explaining 69.47% of the total variance. Good results of the reliability analysis accompanied this ($\alpha = .96$), including both survey content domains, 'Service features for teachers' (SFT) factor ($\alpha = .95$), 'Service features for students' (SFS) factor ($\alpha = .89$) and 'ethics and privacy - teachers' factor ($\alpha = .86$). See Annex Table 10 for details. The factor analysis showed changed factors compared to the factors of the SELAQ student survey. 11 items from the teacher survey showed up in the PCA as the SFT factor. The SF factor, which was transferred in its entirety from the SELAQ to the instructor questionnaire, was retained for the most part. The items 'integrate into feedback' and 'course goals' were dropped. On the other hand, the items 'early interventions' and 'access student data' were newly integrated. The EP items from the SELAQ were not transferred to the teacher survey. However, the factor EP – teachers (EPT) was formed from two newly integrated items. Thus, our survey results analysis delivers sound grounds for interpretations.

### 3.3 Results of Students' Expectations (RQ1)

Regarding RQ1, what the expectations of students for LA at the university are, we found the following results after the analysis of our data. The results of the survey reveal the well-known trend that has been discovered in previous surveys: Items that solicit opinions about the ideal LA implementation typically receive far more support than those that solicit predictions about when the LA implementation will occur [42]. Figures 1 & 2 show the Likert-type item responses of the survey separated by statements regarding 'service features' and 'ethics and privacy' concerns. Figure 1 shows the service features expectations of students (see Annex Table 2 and 3 for details).

Figure 1: Service features (S1–S7) likert-type item responses (7: strongly agree; 1: strongly disagree)

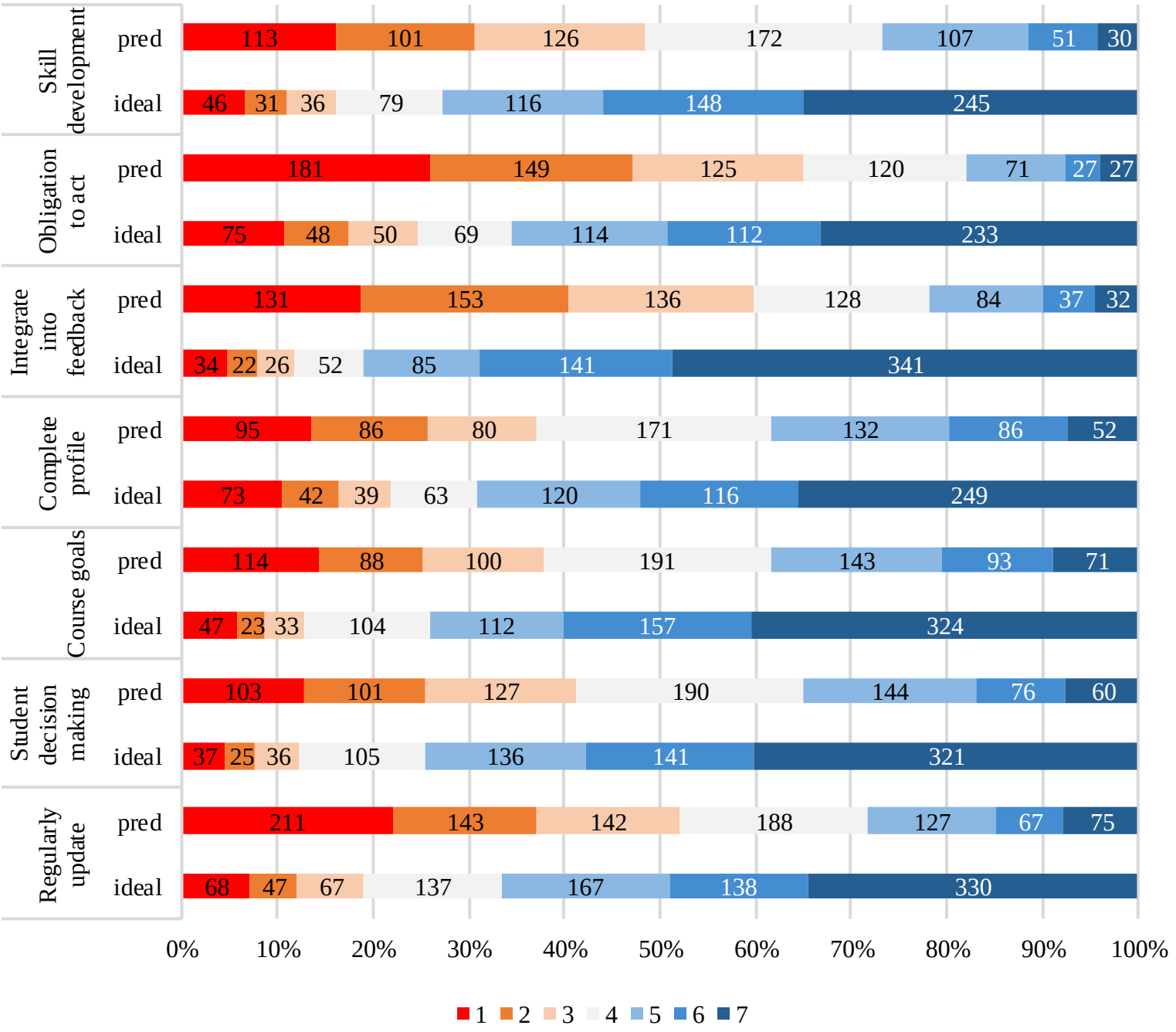

They for instance rated LA as support for 'course goals', 'integration into feedback', 'regular update', 'student decision making' and 'skill development', with high value as a features of LA at the university. Regarding ethics and privacy expectations Figure 2 (see Annex Tables 2 and 3 for details) shows that students rated 'keep data secure', 'identified data', 'third party' transparency, 'consent to collect' and 'alternative purpose' quite high value factors of LA at the university.

Figure 2: Ethics and privacy factor (E1 – E5) likert-type item responses (7: strongly agree; 1: strongly disagree)

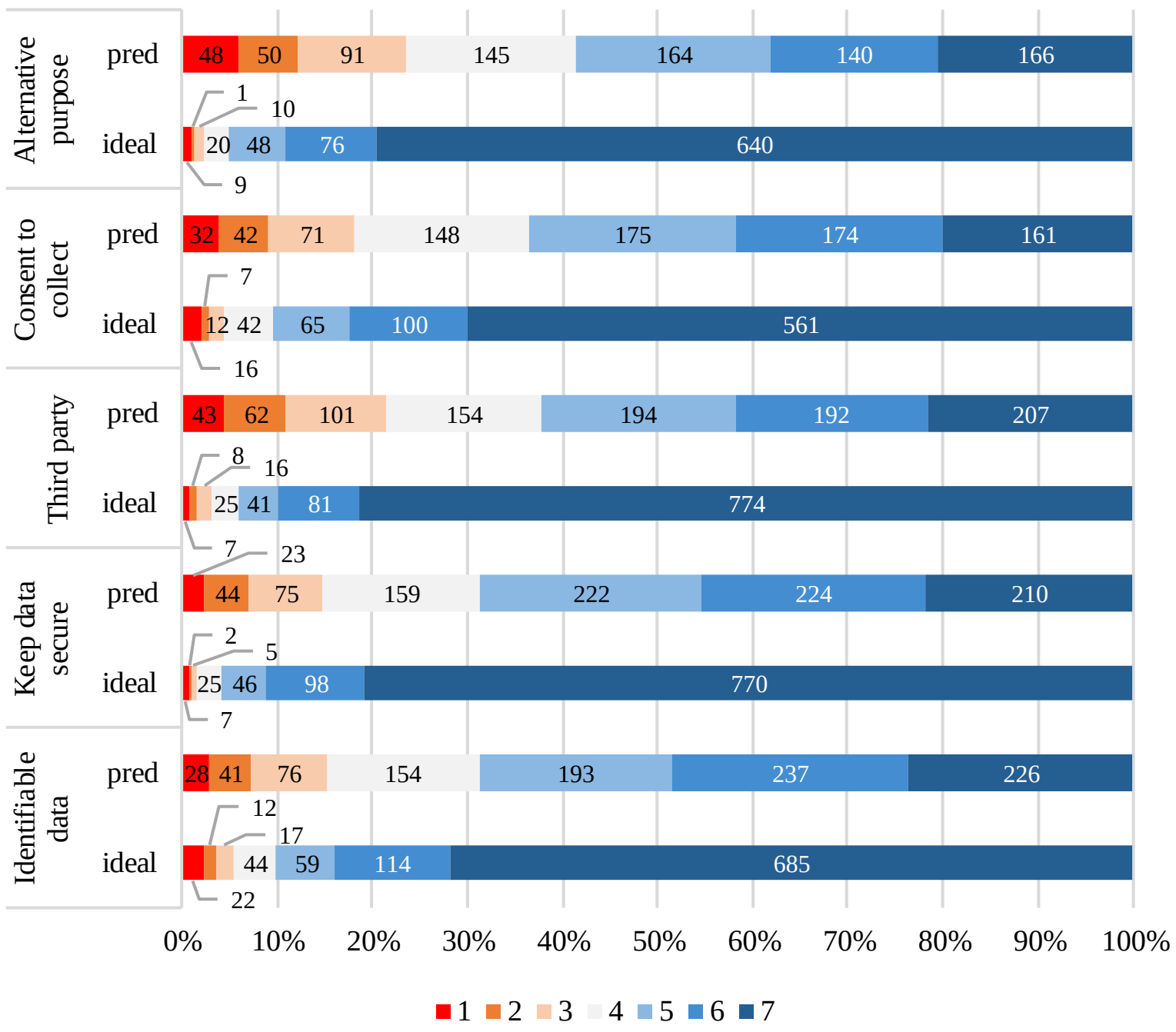

Overall, it becomes clear that there is a high level of agreement, especially with regard to the students' assessment that the involvement of LAs when it takes place ideally. However, the agreement is lower for both factors, service and data protection, when students are asked to assess whether LA involvement will actually happen in this way in reality. More in detail, the results of RQ1 concerning the average of predicted and ideal LA expectations within the student sample are illustrated in Figure 3.

Figure 3: Student survey comparing ideal and realistic estimations towards LA

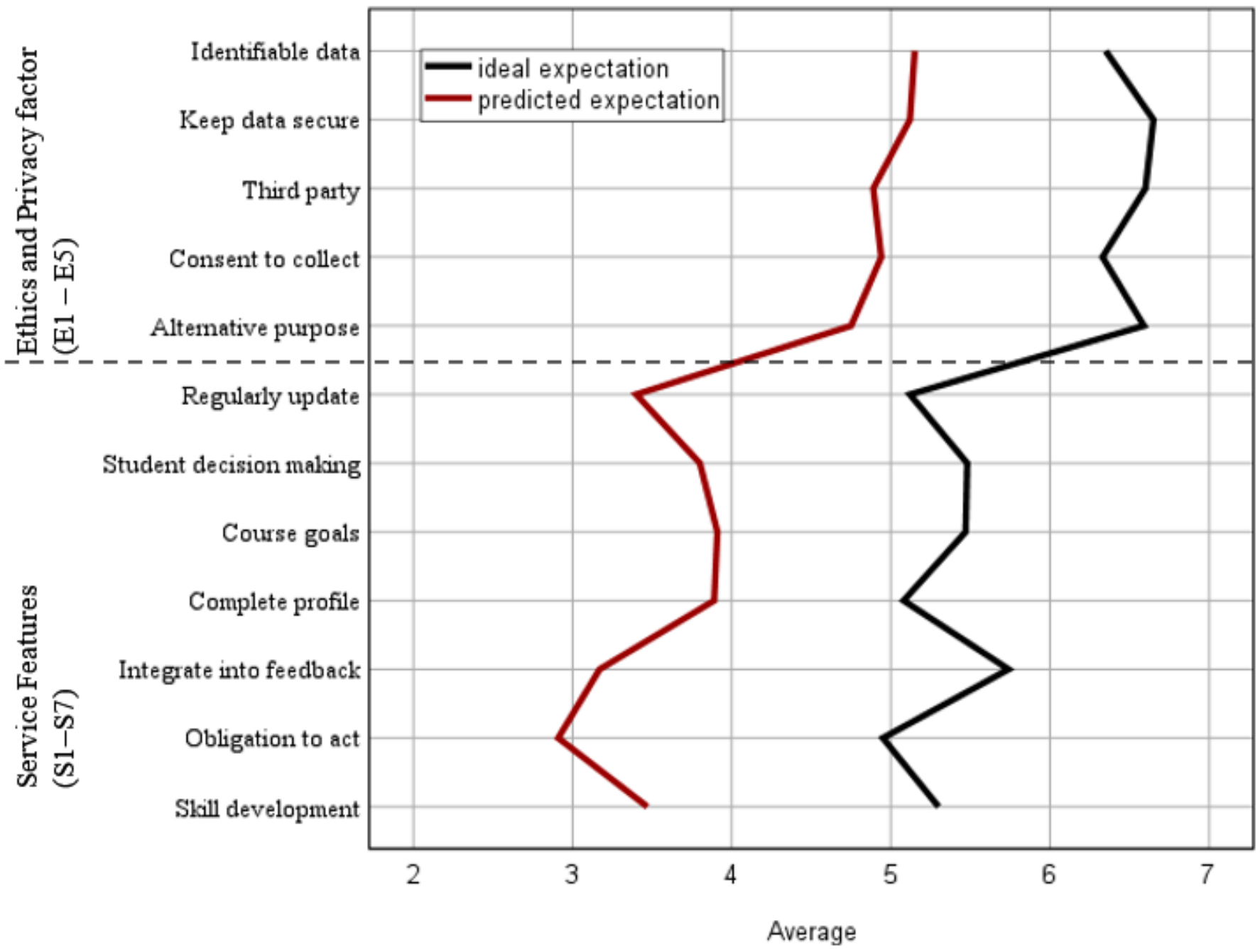

The chart shows that the averaged students' response pattern of expectations on (a) 'service features' (SF) and (b) the 'ethics and privacy factor' (EP) and are quite similar with respect to differences between the items but it differs notably with respect to the degree of agreement, with lower estimations for predicted than for ideal outcomes. The lowest values for students' *predicted expectations* were reported in (a) the service features on obligation to act ($M$ = 2.91), and 'integration into feedback' ($M$ = 3.17). Ratings in (b) EP were overall higher than those in SF but differed only slightly ($4.75 < M < 5.15$). The lowest values within *ideal expectations* of students were found in (a) SF. All of these ratings were notably higher than the ratings in the predicted expectation. The lowest ratings were found for 'obligation to act' ($M$ = 4.95), 'complete profile' ($M$ = 5.08), and 'regularly update' ($M$ = 5.12). Similar to the data on predicted expectations, ratings in (b) EP were overall higher than those in SF but differed only slightly ($6.33 < M < 6.65$). The analysis of *divergence between predicted and ideal expectations* (see Annex Table 11 for details) shows that the lowest divergence in (a) SF is found for 'complete profile' ($d$ = 1.19), and 'course goals' ($d$ = 1.56), and in (b) EP sector for 'identifiable data' ($d$ = 1.20). The highest divergence of ratings in (a) SF was found for 'integrate into feedback' ($d$ = 2.57), which represents by far the highest value for differences between ideal and predicted expectations in the entire data set, and for 'obligation to act' ($d$ = 2.04); in (b) EP, the lowest rating was given for 'alternative purpose' ($d$ = 1.84). To sum up, the fever graph in Figure 3 shows that the values differ between what is expected to happen ideally by the students and what is expected to happen in reality. The ideal case is rated higher than the realistic case for all items, and this difference between the two evaluation modes is the same for almost all items, as shown by the nearly parallel paths of the two lines of the graph. The items concerning EP were rated with higher values on average compared to items concerning SF.

### 3.4 Results of Teachers' Expectations (RQ2)

The results of RQ2 concerning LA expectations of teaching staff are illustrated in Figures 4, 5 and 6 (see Annex Tables 4 and 5 for details).

Figure 4: Service features for teachers (SFT1-SFT11) likert-type item responses (7: strongly agree; 1: strongly disagree).

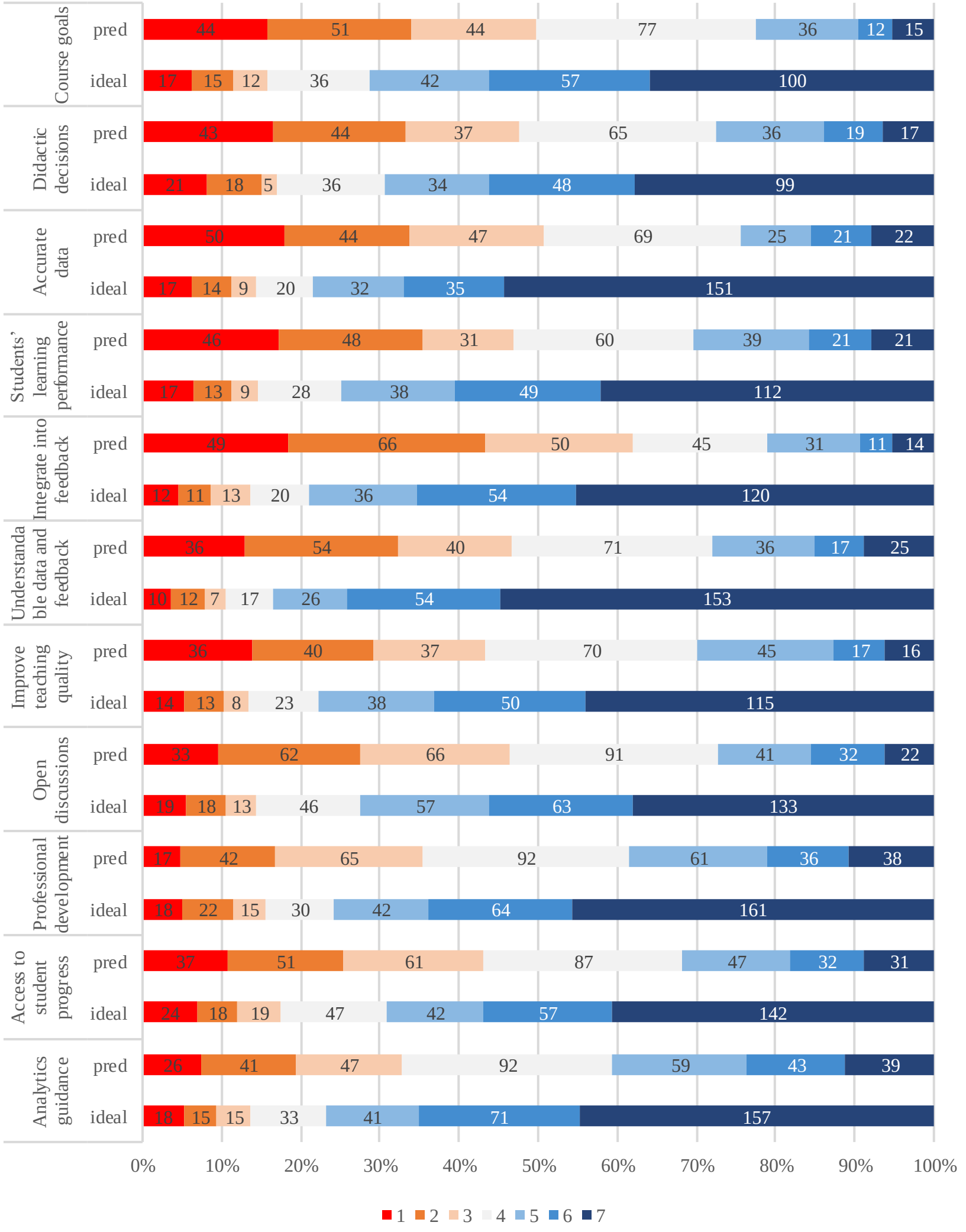

Figure 5: Service features for students (SFS1-SFS7) likert-type item responses
(7: strongly agree; 1: strongly disagree)

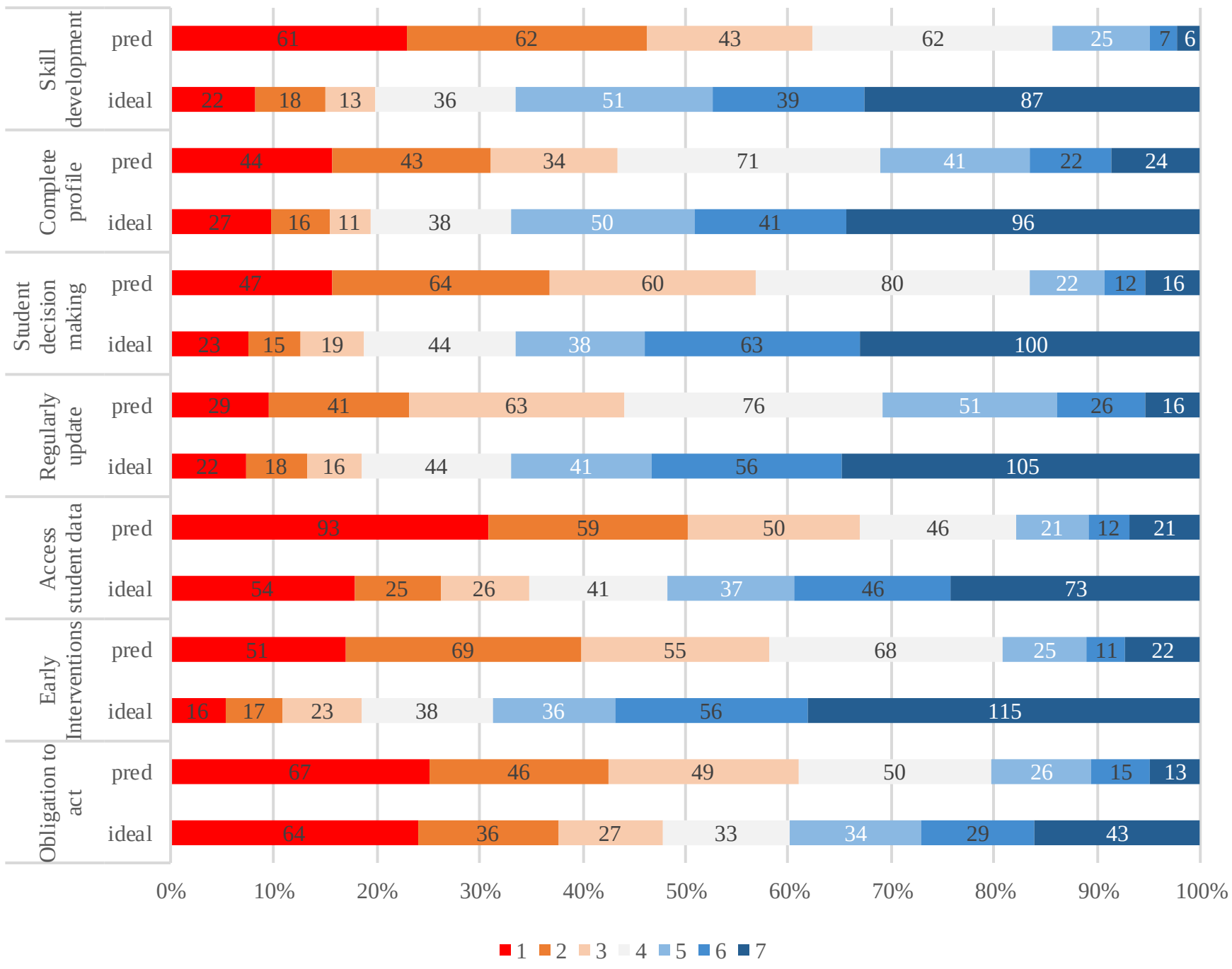

Figure 6: Ethics and privacy -Teacher (EPT1-EPT2) likert-type item responses
(7: strongly agree; 1: strongly disagree)

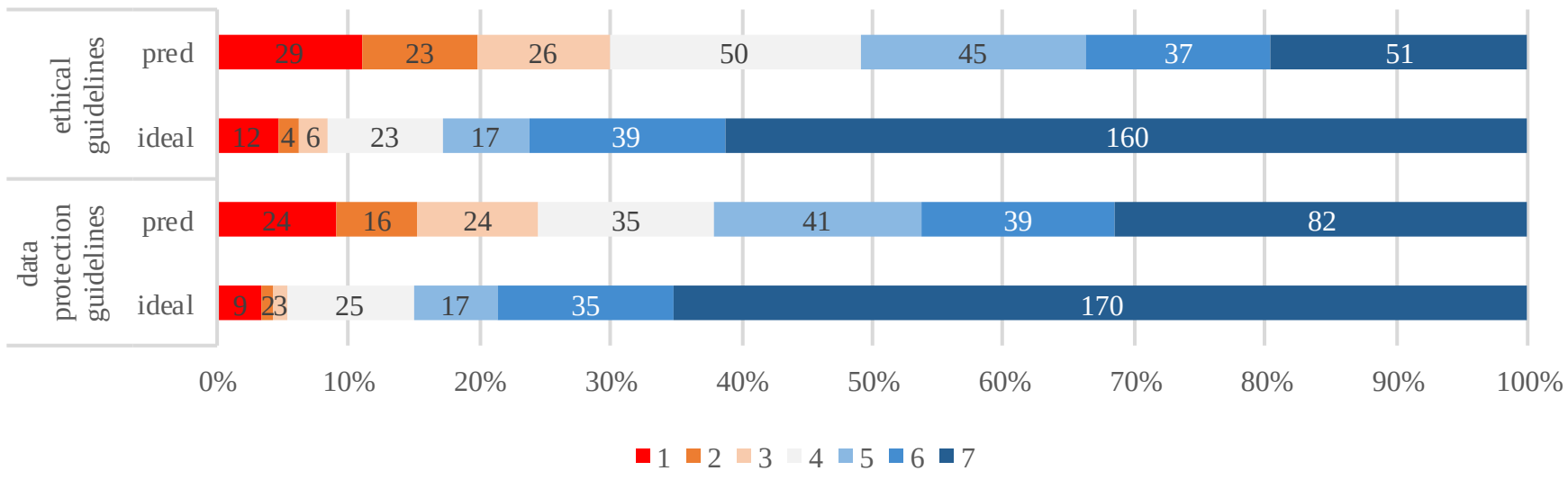

The fever graph in Figure 7 shows that on average, teachers' response patterns of ideal and predicted expectations are rather similar with respect to differences between the items.

Figure 7: Teachers survey comparing ideal and realistic estimations towards LA (see annex table 2 for details)

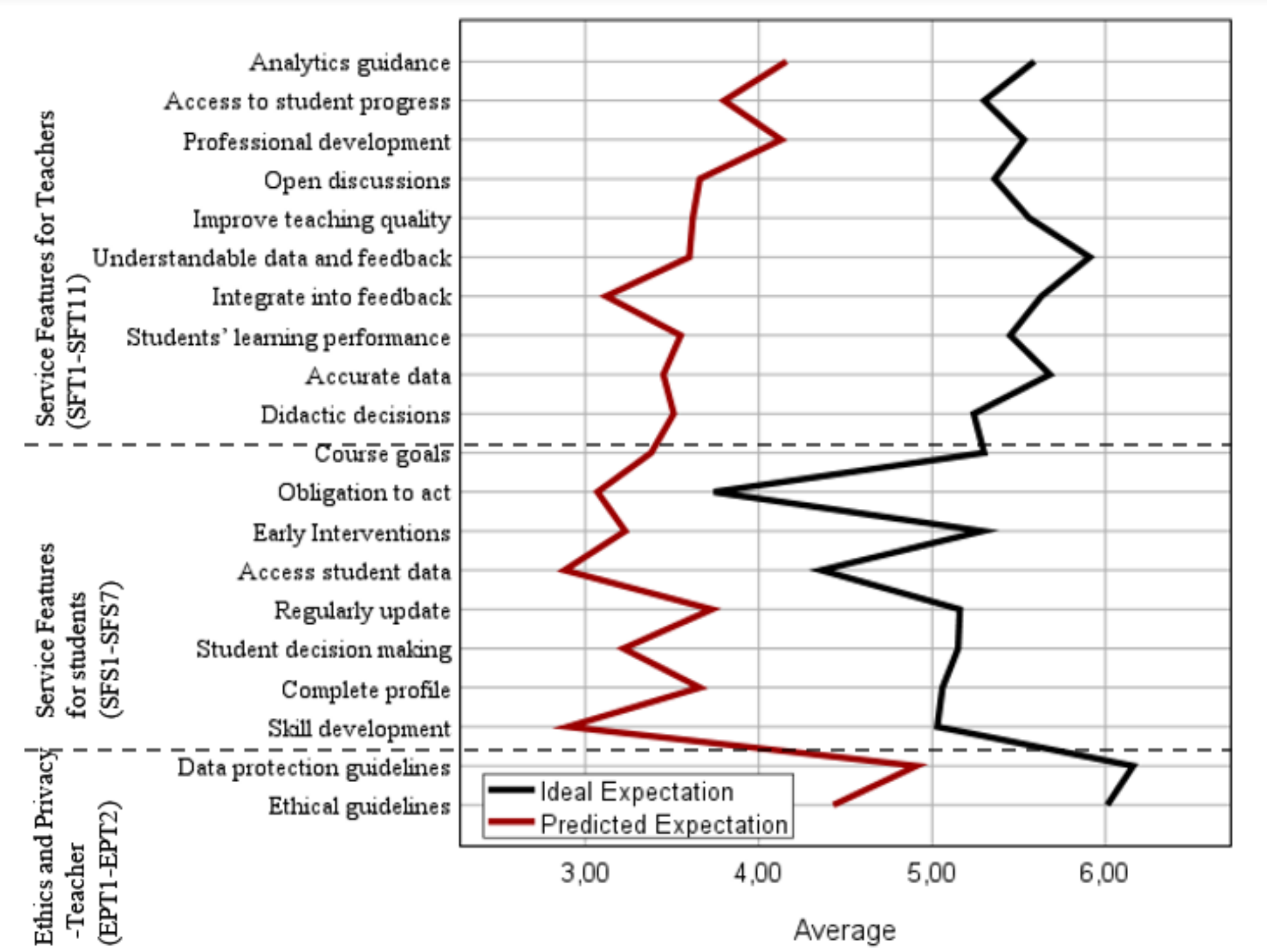

Expectations differ considerably with respect to the degree of agreement, though, with lower estimations for predicted as opposed to ideal outcomes. In detail, for *predicted expectations* the lowest ratings were reported in (a) SFS for 'access student data' (*M* = 2.88), 'skill development' (*M* = 2.90), and 'obligation to act (*M* = 3.07*)*. The lowest rating in (b) SFT was reported for 'integrate into feedback' (*M* = 3.12). Ratings in (c) EPT differed only slightly. The by far highest rating within predicted expectations of teachers was found in (c) EPT for 'data protection guidelines' (*M* = 4.91). Further rather high ratings were found in (b) SFT for 'analytics guidance' (*M* = 4.16) and 'professional development' (*M* = 4.13). In the results on teachers' *ideal expectations*, the by far lowest rating on was found in (a) SFS for 'obligation to act' (*M* = 3.74). Ratings in (b) SFT did not differ remarkably. The highest values are found in (c) EPT, with 'data protection' (*M* = 6.16) representing the item with the highest expectation rating in the averaged data sample of teachers. The analysis of *divergence between predicted and ideal expectations* (see Annex Table 12 for details) shows that the lowest divergence is found in (a) SFS for 'obligation to act' (*d* = 0.67), which represents the lowest divergence of all items explored. The highest divergence values were found in (b) SFT, for 'integrate into feedback' (*d* = 2.51), 'understandable data and feedback (*d* = 2.31), and 'accurate data' (*d* = 2.23).

### 3.5 Results of Comparison between Teachers and Students (RQ3)

In a last step, we compared the results of the students and teacher survey items (RQ3) that were identical and looked for differences and communalities: In comparing teachers' and students' expectations, only seven items regarding service features (S1-S7) of LA were identically in both questionnaires and, thus, allow comparative descriptive analyses. Figure 8 shows the direct comparison in a fever graph.

Figure 8: Comparison of teachers & students estimations towards LA

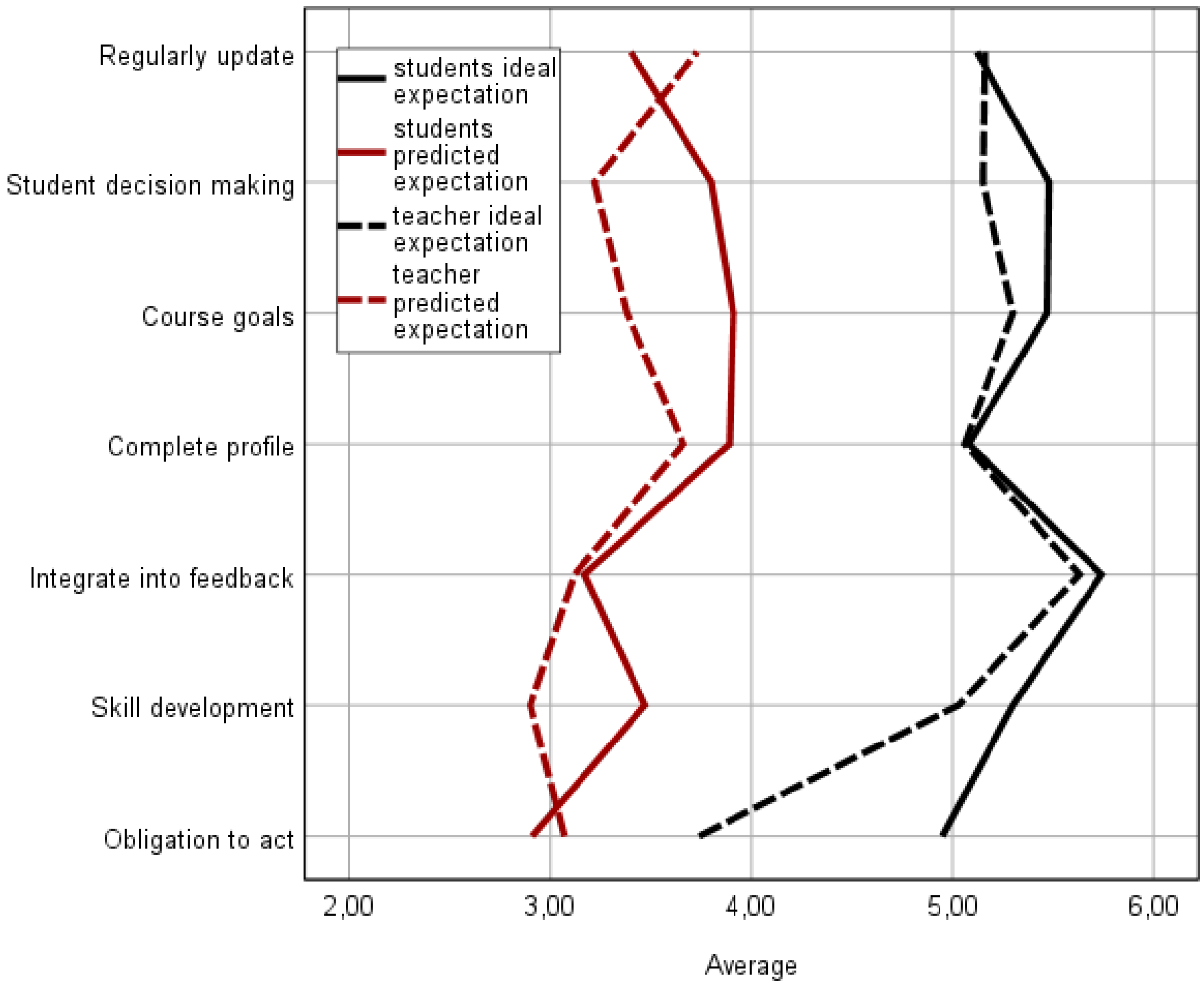

It is noteworthy that the items 'student decision-making' (*ds* = 1.68; *dt* =1.93) and 'course goal' (*ds* = 1.56; *dt* = 1.92) were valued higher by the students in the ideal and realistic case than by teachers overall. This was also the case for 'skill development' (*ds* = 2.04, *dt* = 0.67). The question regarding 'a complete profile' was rated closest to being ideal and realistic to happen by both teachers and students. This was different for the item 'obligation to act'. There was a clear difference between teachers (ideal: 3.74, pred. 3.07; diff: 0.67) and students (ideal: 4.95, pred.: 2.91; diff: 2.04). In the ideal case, teachers rated the obligation to act for teaching staff about implementing LA lower than students. In the realistic case, teachers and students rated it similarly low to become a duty in real life (see Annex Table 13 for details).

## 4 DISCUSSION

Concluding from our analysis, we found that students (RQ1) generally shared a high level of agreement with regard to the importance of LAs when it takes place in an ideal way. However, the agreement was lower for both service and data protection factors when students were asked to assess whether LA involvement would actually happen in this way in reality. Assessing the expectations of teachers (RQ2), our results were similar. Their ideal and predicted expectations towards service features for students, teachers and ethics and privacy also differed between the items for students. However, besides many commonalities between students and teachers, there were also some noteworthy differences (RQ3). In detail, in direct comparison, teachers valued service features like 'student decision-making' and 'course goal' lower than students, ideally and expected. Especially regarding the 'obligation to act', the expectations differed widely. Teachers found the duty to act in the implementation of LA at the HEI of the lowest value overall in the survey. This differs quite remarkably from the students' ratings, who found it important ideally and realistically. It was the second-highest expected value in the student survey. The difference indicated high expectation differences for teachers to act upon LA ideally; however, teachers and students are both agreeing on the limited readiness of teachers to act in this regard quite realistically. In the literature, it is highlighted that the desire of students for the faculty to take responsibility to act on student data provided by LA should be addressed. Our results support warnings by Whitelock-Wainwright et al. (2021) that if faculty disregards this aspect when LA is introduced, the student's willingness to cooperate and provide data might suffer. Moreover, our findings align with a recent European literature review by Santos et al. (2022) which highlights that while HEI professors are generally aware of available digital tools, their utilization remains limited. The review underscores the importance of teacher training, emphasizing the need for a thorough assessment of professors' digital competence levels as a prerequisite for personalized training initiatives. Interesting was also the expectation towards 'a complete profile' [41]. There were little deviations between students and teachers, and both valued the desired and expected outcome of implementing a complete technical dashboard profile at the university for realistic. This falls into place with other findings about our university. Biedermann et al. (2019) found in their survey on administrative staff, teachers and students at the university, that wishes for hardware or service do exist but are seen as very feasible to solve and not very important for a successful digitalization strategy [2]. The study's results rather showed that there was an agreement between all groups that matters of culture and mentality remain the biggest factors for the successful digitalization in HEI. Furthermore, this evaluation was ranked highest among the group of employees. One can assume that this group has experienced difficulties in achieving cultural changes. Our findings indicate that technical and especially human-related factors play an essential role in technology acceptance processes, with theoretical and practical implications. Human factors should be considered in further research to investigate antecedents of user acceptance concerning LA and AI in HEI. Concerning practical recommendations, our results yield the relevance of developing and implementing institutional engagement strategies to nurture the acceptance of LA in the HEI and support the practical rollout of LA and AI-supported educational technology in higher education.

## 5 LIMITATIONS

This paper has several limitations. First, we chose a cross-sectional study based on self-reports, not a longitudinal one, which could have yielded more knowledge about the estimations and variations over time. Second, our instruments were not congruent in their design, so the surveys of teachers and students were not comparable in all their items. Also, since they were translated from English to German, the information we collected might vary slightly in German from the original ones. Third, the survey items referred to LA and did not assess whether participants' attitudes towards LA differ from those towards AI. Fourth, the sample included only students and teachers from our university, not surveying further stakeholders such as managers and administrative staff, whose opinion is also crucial to consider when an institutional-wide strategy is to be developed. Lastly, we cannot draw comparisons to other institutions in our study since we decided to only collect data from our HEI in line with the internal strategy development.

## 6 CONCLUSION

Stakeholder attitudes have been highlighted to be an important determinant for the success of an implemented service such as LA, which includes not only a flawless technical rollout and maintenance but also user acceptance as a prerequisite for the actual use of tools [11]. In this paper, the authors investigated students and teaching staff at a University in Germany by using quantitative surveys derived from the SHEILA framework [31, 35]. Explorative analyses allowed insights and comparisons of two stakeholder groups within one HEI. Overall, empirical findings showed that the predicted expectations were notably lower in both groups than the ideal expectations. Regarding differences between the groups, the most salient aspect was that students reported a strong desire for faculty members' obligation to be trained and act on LA data. In contrast, teachers did not agree with the idea of such mandatory accountability. The realistic expectations of both groups met halfway, though, indicating that they share visions of realistic LA scenarios at their HEI. Our results yield the practical implication that the stakeholders' expectations should be integrated into a targeted stakeholder engagement strategy at HEIs that supports HEIs in overcoming the social challenges of implementing LA in higher education.

## 7 BIBLIOGRAPHY

[1] Baker, Ryan S. (2019): Challenges for the Future of Educational Data Mining: The Baker Learning Analytics Prizes. In: *Journal of Educational Data Mining* 11 (1), S. 1–17. DOI: 10.5281/ZENODO.3554745.

[2] Biedermann, Daniel; Kalbfell, Lea; Schneider, Jan; Drachsler, Hendrik (2019): Stakeholder attitudes towards digitalization in higher education institutions. In: Niels Pinkwart und Johannes Konert (Hg.): DELFI 2019. Bonn: Gesellschaft für Informatik e.V, S. 57–66.

[3] Bond, Melissa; Marín, Victoria I.; Dolch, Carina; Bedenlier, Svenja; Zawacki-Richter, Olaf (2018): Digital transformation in German higher education: student and teacher perceptions and usage of digital media. In: *Int J Educ Technol High Educ* 15 (1). DOI: 10.1186/s41239-018-0130-1.

[4] Ciordas-Hertel, George-Petru; Schneider, Jan; Ternier, Stefaan; Drachsler, Hendrik (2019): Adopting trust in learning analytics infrastructure. A structured literature review. DOI: 10.25656/01:23312.

[5] Drachsler, Hendrik; Greller, Wolfgang (2016): Privacy and analytics. In: Dragan Gašević, Grace Lynch, Shane Dawson, Hendrik Drachsler und Carolyn Penstein Rosé (Hg.): Proceedings of the Sixth International Conference on Learning Analytics & Knowledge - LAK '16. the Sixth International Conference. Edinburgh, United Kingdom, 25.04.2016 - 29.04.2016. New York, New York, USA: ACM Press, S. 89–98.

[6] Drachsler, Hendrik; Goldhammer, Frank (2020): Learning Analytics and eAssessment—Towards Computational Psychometrics by Combining Psychometrics with Learning Analytics. In: Daniel Burgos (Hg.): Radical Solutions and Learning Analytics. Personalised Learning and Teaching Through Big Data. 1st ed. 2020. Singapore: Springer Singapore; Imprint Springer (Lecture Notes in Educational Technology), S. 67–80.

[7] Falcão, Taciana Pontual; Mello, Rafael Ferreira; Rodrigues, Rodrigo Lins; Diniz, Juliana Regueira Basto; Tsai, Yi-Shan; Gašević, Dragan (2020): Perceptions and expectations about learning analytics from a brazilian higher education institution. In: Christoph Rensing, Hendrik Drachsler, Vitomir Kovanović, Niels Pinkwart, Maren Scheffel und Katrien Verbert (Hg.): Proceedings of the Tenth International Conference on Learning Analytics & Knowledge. LAK '20: 10th International Conference on Learning Analytics and Knowledge. Frankfurt Germany, 23 03 2020 27 03 2020. New York, NY, USA: ACM, S. 240–249.

[8] Field, Andy (2018): Discovering statistics using IBM SPSS statistics. 5th edition. Los Angeles, London, New Delhi, Singapore, Washington DC, Melbourne: SAGE (SAGE edge).

[9] Farrell, Tracie; Mikroyannidis, Alexander; Alani, Harith (2017): "We're Seeking Relevance": Qualitative Perspectives on the Impact of Learning Analytics on Teaching and Learning. In: Élise Lavoué, Hendrik Drachsler, Katrien Verbert, Julien Broisin und Mar Pérez-Sanagustín (Hg.): Data Driven Approaches in Digital Education, Bd. 10474. Cham: Springer International Publishing (Lecture Notes in Computer Science), S. 397–402.

[10] Francis, Peter; Broughan, Christine; Foster, Carly; Wilson, Caroline (2020): Thinking critically about learning analytics, student outcomes, and equity of attainment. In: *Assessment & Evaluation in Higher Education* 45 (6), S. 811–821. DOI: 10.1080/02602938.2019.1691975.

[11] Greller, Wolfgang; Drachsler, Hendrik (2012): Translating Learning into Numbers: A Generic Framework for Learning Analytics. In: *Journal of Educational Technology & Society* 15 (3), S. 42–57. Online http://www.jstor.org/stable/jeductechsoci.15.3.42.

[12] Gray, Geraldine; Schalk, Ana Elena; Cooke, Gordon; Murnion, Phelim; Rooney, Pauline; O'Rourke, K. C. (2022): Stakeholders' insights on learning analytics: Perspectives of students and staff. In: *Computers & Education* 187, S. 104550. DOI: 10.1016/j.compedu.2022.104550.

[13] Gašević, Dragan; Tsai, Yi-Shan; Drachsler, Hendrik (2022): Learning analytics in higher education – Stakeholders, strategy and scale. In: *The Internet and Higher Education* 52, S. 100833. DOI: 10.1016/j.iheduc.2021.100833.

[14] Hilliger, Isabel; Ortiz-Rojas, Margarita; Pesántez-Cabrera, Paola; Scheihing, Eliana; Tsai, Yi-Shan; Muñoz-Merino, Pedro J. et al. (2020): Identifying needs for learning analytics adoption in Latin American universities: A mixed-methods approach. In: *The Internet and Higher Education* 45, S. 100726. DOI: 10.1016/j.iheduc.2020.100726.

[15] Hamet, Pavel; Tremblay, Johanne (2017): Artificial intelligence in medicine. In: *Metabolism - Clinical and Experimental* 69S, S36-S40. DOI: 10.1016/j.metabol.2017.01.011.

[16] Ifenthaler, Dirk; Schumacher, Clara (2016): Student perceptions of privacy principles for learning analytics. In: *Education Tech Research Dev* 64 (5), S. 923–938. DOI: 10.1007/s11423-016-9477-y.

[17] Ifenthaler, Dirk; Yau, Jane Yin-Kim (2019): Higher Education Stakeholders' Views on Learning Analytics Policy Recommendations for Supporting Study Success. In: *iJAI* 1 (1), S. 28. DOI: 10.3991/ijai.v1i1.10978.

[18] Jones, Kyle M. L.; Asher, Andrew; Goben, Abigail; Perry, Michael R.; Salo, Dorothea; Briney, Kristin A.; Robertshaw, M. Brooke (2020): "We're being tracked at all times": Student perspectives of their privacy in relation to learning analytics in higher education. In: *Journal of the Association for Information Science and Technology* 71 (9), S. 1044–1059. DOI: 10.1002/asi.24358.

[19] Knight, Simon; Gibson, Andrew; Shibani, Antonette (2020): Implementing learning analytics for learning impact: Taking tools to task. In: *The Internet and Higher Education* 45, S. 100729. DOI: 10.1016/j.iheduc.2020.100729.

[20] Khan, Osama (2017): Learners' and Teachers' Perceptions of Learning Analytics (LA): A Case Study of Southampton Solent University (SSU). In: *International Association for Development of the Information Society*. Online verfügbar unter https://eric.ed.gov/?id=ed579477.

[21] Klein, Carrie; Lester, Jaime; Rangwala, Huzefa; Johri, Aditya (2019): Learning Analytics Tools in Higher Education: Adoption at the Intersection of Institutional Commitment and Individual Action. In: *The Review of Higher Education* 42 (2), S. 565–593. DOI: 10.1353/rhe.2019.0007.

[22] Kollom, Kaire; Tammets, Kairit; Scheffel, Maren; Tsai, Yi-Shan; Jivet, Ioana; Muñoz-Merino, Pedro J. et al. (2021): A four-country cross-case analysis of academic staff expectations about learning analytics in higher education. In: *The Internet and Higher Education* 49, S. 100788. DOI: 10.1016/j.iheduc.2020.100788.

[23] Long, Phil; Siemens, George (2011): Penetrating the Fog: Analytics in Learning and Education. In: *EDUCAUSE Review* 46 (5), S. 30–40. Online verfügbar unter https://er.educause.edu/articles/2011/9/penetrating-the-fog-analytics-in-learning-and-education.

[24] Mahmoud, Mai; Dafoulas, Georgios; Abd ElAziz, Rasha; Saleeb, Noha (2020): Learning analytics stakeholders' expectations in higher education institutions: a literature review. In: *IJILT* 38 (1), S. 33–48. DOI: 10.1108/IJILT-05-2020-0081.

[25] Mercader, Cristina, Gairín, Joaquín. 2020. University teachers' perception of barriers to the use of digital technologies: the importance of the academic discipline. *International Journal of Educational Technology in Higher Education*, *17*(1), 4. https://doi.org/10.1186/s41239-020-0182-x

[26] Okkonen, Jussi; Helle, Tanja; Lindsten, Hanna (2020): Expectation Differences Between Students and Staff of Using Learning Analytics in Finnish Universities. In: Álvaro Rocha, Carlos Ferrás, Carlos Enrique Montenegro Marin und Víctor Hugo Medina García (Hg.): Information Technology and Systems, Bd. 1137. Cham: Springer International Publishing (Advances in Intelligent Systems and Computing), S. 383–393.

[27] Prieto, Luis P.; Rodríguez-Triana, María Jesús; Martínez-Maldonado, Roberto; Dimitriadis, Yannis; Gašević, Dragan (2019): Orchestrating learning analytics (OrLA): Supporting inter-stakeholder communication about adoption of learning analytics at the classroom level. In: *AJET* 35 (4). DOI: 10.14742/ajet.4314.

[28] Roberts, Lynne D.; Howell, Joel A.; Seaman, Kristen; Gibson, David C. (2016): Student Attitudes toward Learning Analytics in Higher Education: "The Fitbit Version of the Learning World". In: *Frontiers in psychology* 7, S. 1959. DOI: 10.3389/fpsyg.2016.01959.

[29] Santos, Cassio, Pedro, Nueza, Mattar, João. 2022. Digital Competence of Higher Education Professors in the European Context: A Scoping Review Study. *International Journal of Emerging Technologies in Learning (iJET)*, *17*(18), 222–242. https://doi.org/10.3991/ijet.v17i18.31395

[30] Stelmaszak, Marta; Aaltonen, Aleksi (2018): Closing the Loop of Big Data Analytics: the Case of Learning Analytics. Research Papers (ECIS 2018 Proceedings, 82). Online verfügbar unter https://aisel.aisnet.org/ecis2018_rp/82, zuletzt geprüft am 27.03.2023.

[31] Scheffel, Maren; Drachsler, Hendrik; Toisoul, Christian; Ternier, Stefaan; Specht, Marcus (2017): The Proof of the Pudding: Examining Validity and Reliability of the Evaluation Framework for Learning Analytics. In: Élise Lavoué, Hendrik Drachsler, Katrien Verbert, Julien Broisin und Mar Pérez-Sanagustín (Hg.): Data Driven Approaches in Digital Education, Bd. 10474. Cham: Springer International Publishing (Lecture Notes in Computer Science), S. 194–208.

[32] Schneckenberg, Dirk (2009): Understanding the real barriers to technology-enhanced innovation in higher education. In: *Educational Research* 51 (4), S. 411–424. DOI: 10.1080/00131880903354741.

[33] Schumacher, Clara; Ifenthaler, Dirk (2018): Features students really expect from learning analytics. In: *Computers in Human Behavior* 78, S. 397–407. DOI: 10.1016/j.chb.2017.06.030.

[34] Silvola, Anni; Näykki, Piia; Kaveri, Anceli; Muukkonen, Hanni (2021): Expectations for supporting student engagement with learning analytics: An academic path perspective. In: *Computers & Education* 168, S. 104192. DOI: 10.1016/j.compedu.2021.104192.

[35] Tsai, Yi-Shan; Gašević, Dragan; Whitelock-Wainwright, Alexander; Muñoz-Merino, Pedro J.; Moreno-Marcos, Pedro M.; Fernández, Aarón Rubio et al. (2018): SHEILA. Research Report November 2018. Online: sheilaproject.eu.

[36] Tsai, Yi-Shan; Moreno-Marcos, Pedro Manuel; Jivet, Ioana; Scheffel, Maren; Tammets, Kairit; Kollom, Kaire; Gašević, Dragan (2018): The SHEILA Framework: Informing Institutional Strategies and Policy Processes of Learning Analytics. In: *Learning Analytics* 5 (3). DOI: 10.18608/jla.2018.53.2.

[37] Thompson, A. G.; Suñol, R. (1995): Expectations as determinants of patient satisfaction: concepts, theory and evidence. In: *International journal for quality in health care : journal of the International Society for Quality in Health Care* 7 (2), S. 127–141. DOI: 10.1093/intqhc/7.2.127.

[38] Tsai, Yi-Shan; Whitelock-Wainwright, Alexander; Gašević, Dragan (2020): The privacy paradox and its implications for learning analytics. In: Christoph Rensing, Hendrik Drachsler, Vitomir Kovanović, Niels Pinkwart, Maren Scheffel und Katrien Verbert (Hg.): Proceedings of the Tenth International Conference on Learning Analytics & Knowledge. LAK '20: 10th International Conference on Learning Analytics and Knowledge. Frankfurt Germany, 23 03 2020 27 03 2020. New York, NY, USA: ACM, S. 230–239.

[39] West, Deborah; Huijser, Henk; Heath, David; Lizzio, Alf; Toohey, Danny; Miles, Carol et al. (2015): Higher Education Teachers’ Experiences with Learning Analytics in Relation to Student Retention. In: *AJET*. DOI: 10.14742/ajet.3435.

[40] Whitelock-Wainwright, Alexander; Gašević, Dragan; Tsai, Yi-Shan; Drachsler, Hendrik; Scheffel, Maren; Muñoz-Merino, Pedro J. et al. (2020): Assessing the validity of a learning analytics expectation instrument: A multinational study. In: *J Comput Assist Learn* 36 (2), S. 209–240. DOI: 10.1111/jcal.12401.

[41] Whitelock-Wainwright, Alexander; Tsai, Yi-Shan; Drachsler, Hendrik; Scheffel, Maren; Gašević, Dragan (2021): An exploratory latent class analysis of student expectations towards learning analytics services. In: *The Internet and Higher Education* 51, S. 100818. DOI: 10.1016/j.iheduc.2021.100818.

[42] Wollny, Sebastian; Di Mitri, Daniele; Jivet, Ioana; Muñoz-Merino, Pedro; Scheffel, Maren; Schneider, Jan et al. (2023): Students' expectations of Learning Analytics across Europe. In: *Computer Assisted Learning*, Artikel jcal.12802. DOI: 10.1111/jcal.12802.

# 8 Annex

Table 2: Values and correlation of the students survey for the ideally expectations ('Ideally, I would like this to happen')

| | | *n* | *M* | *SD* | 1 | 2 | 3 | 4 | 5 | 6 | 7 | 8 | 9 | 10 | 11 | 12 |
|---|---|---|---|---|---|---|---|---|---|---|---|---|---|---|---|---|
| Ethics and Privacy factor (E1–E5) | Identifiable data | 953 | 6.35 | 1.337 | -- | | | | | | | | | | | |
| | Keep data secure | 953 | 6.65 | .899 | .443** | -- | | | | | | | | | | |
| | Third party | 952 | 6.60 | 1.037 | .459** | .535** | -- | | | | | | | | | |
| | Consent to collect | 803 | 6.3 | 1.285 | .502** | .465** | .556** | -- | | | | | | | | |
| | Alternative purpose | 804 | 6.59 | 1.011 | .394** | .520** | .551** | .636** | -- | | | | | | | |
| Service Features factor (S1-S7) | Regularly update | 954 | 5.12 | 1.878 | .119** | .153** | .155** | .166** | .136** | -- | | | | | | |
| | Student decision making | 801 | 5.48 | 1.698 | .064 | .094** | .081* | .128** | .163** | .560** | -- | | | | | |
| | Course goals | 800 | 5.47 | 1.758 | .038 | .073* | .115** | .139** | .182** | .564** | .761** | -- | | | | |
| | Complete profile | 702 | 5.08 | 2.024 | .003 | -.014 | .015 | .018 | -.009 | .527** | .637** | .658** | -- | | | |
| | Integrate into feedback | 701 | 5.74 | 1.699 | .047 | .095* | .080* | .112** | .064 | .389** | .536** | .519** | .580** | -- | | |
| | Obligation to act | 701 | 4.95 | 2.051 | -.042 | -.012 | .016 | .023 | .011 | .454** | .496** | .462** | .502** | .533** | -- | |
| | Skill development | 701 | 5.30 | 1.814 | .021 | .054 | .017 | .077* | .040 | .454** | .582** | .566** | .551** | .612** | .600** | -- |

** The correlation is significant at the 0.01 level (2-sided). * The correlation is significant at the 0.05 level (2-sided)

Table 3: Values and correlation of the students survey for the predicted expectations ('I expect this to happen in reality')

| | | *n* | *M* | *SD* | 1 | 2 | 3 | 4 | 5 | 6 | 7 | 8 | 9 | 10 | 11 | 12 |
|---|---|---|---|---|---|---|---|---|---|---|---|---|---|---|---|---|
| Ethics and Privacy factor (E1–E5) | Identifiable data | 955 | 5.15 | 1.584 | -- | | | | | | | | | | | |
| | Keep data secure | 957 | 5.12 | 1.542 | .514** | -- | | | | | | | | | | |
| | Third party | 953 | 4.89 | 1.719 | .511** | .587** | -- | | | | | | | | | |
| | Consent to collect | 803 | 4.94 | 1.632 | .631** | .506** | .569** | -- | | | | | | | | |
| | Alternative purpose | 804 | 4.75 | 1.758 | .557** | .558** | .628** | .693** | -- | | | | | | | |
| Service Features factor (S1-S7) | Regularly update | 953 | 3.40 | 1.872 | .273** | .213** | .273** | .361** | .322** | -- | | | | | | |
| | Student decision making | 801 | 3.80 | 1.737 | .267** | .235** | .262** | .357** | .347** | .552** | -- | | | | | |
| | Course goals | 800 | 3.91 | 1.812 | .216** | .180** | .227** | .322** | .305** | .575** | .748** | -- | | | | |
| | Complete profile | 702 | 3.89 | 1.777 | .236** | .188** | .239** | .333** | .271** | .500** | .571** | .678** | -- | | | |
| | Integrate into feedback | 701 | 3.17 | 1.676 | .216** | .262** | .306** | .297** | .334** | .437** | .514** | .505** | .555** | -- | | |
| | Obligation to act | 700 | 2.91 | 1.669 | .146** | .179** | .230** | .209** | .260** | .446** | .466** | .452** | .508** | .638** | -- | |
| | Skill development | 700 | 3.47 | 1.664 | .236** | .283** | .301** | .305** | .344** | .443** | .520** | .551** | .551** | .596** | .622** | -- |

** The correlation is significant at the 0.01 level (2-sided). * The correlation is significant at the 0.05 level (2-sided)

Table 4: Values and correlation of the teacher survey for the ideally expectations ('Ideally, I would like this to happen')

| | | *n* | *M* | *SD* | 1 | 2 | 3 | 4 | 5 | 6 | 7 | 8 | 9 | 10 | 11 | 12 | 13 | 14 | 15 | 16 | 17 | 18 | 19 | 20 |
|---|---|---|---|---|---|---|---|---|---|---|---|---|---|---|---|---|---|---|---|---|---|---|---|---|
| Service Features for | Analytics guidance | 350 | 5.59 | 1.771 | -- | | | | | | | | | | | | | | | | | | | |
| | Access to student progress | 349 | 5.30 | 1.904 | .747** | -- | | | | | | | | | | | | | | | | | | |

| | | | | | | | | | | | | | | | | | | | | | | | |
|---|---|---|---|---|---|---|---|---|---|---|---|---|---|---|---|---|---|---|---|---|---|---|---|
| | Professional development | 352 | 5.53 | 1.836 | .802** | .719* * | -- | | | | | | | | | | | | | | | | |
| | Open discussions | 349 | 5.36 | 1.788 | .666** | .660** | .759** | -- | | | | | | | | | | | | | | | |
| | Improve teaching quality | 261 | 5.56 | 1.781 | .641** | .623** | .657** | .635** | -- | | | | | | | | | | | | | | |
| | Understandable data and feedback | 279 | 5.91 | 1.642 | .595** | .535** | .592** | .543** | .632** | -- | | | | | | | | | | | | | |
| | Integrate into feedback | 266 | 5.63 | 1.729 | .604** | .569** | .624** | .574** | .729** | .646** | -- | | | | | | | | | | | | |
| | Students' learning performance | 266 | 5.45 | 1.839 | .641** | .572** | .626** | .575** | .828** | .635** | .739** | -- | | | | | | | | | | | |
| | Accurate data | 278 | 5.68 | 1.866 | .568** | .530** | .560** | .553** | .704** | .688** | .723** | .711** | -- | | | | | | | | | | |
| | Didactic decisions | 261 | 5.24 | 1.945 | .571** | .563** | .618** | .622** | .872** | .593** | .643** | .797** | .633** | -- | | | | | | | | | |
| | Course goals | 279 | 5.30 | 1.822 | .512** | .568** | .549** | .542** | .728** | .617** | .618** | .670** | .655** | .644** | -- | | | | | | | | |
| Service Features for students (SFS1- FS7) | Obligation to act | 266 | 3.74 | 2.185 | .260** | .282** | .421** | .339** | .402** | .291** | .374** | .430** | .314** | .432** | .384** | -- | | | | | | | |
| | Early interventions | 301 | 5.29 | 1.853 | .413** | .443** | .494** | .456** | .601** | .471** | .547** | .611** | .600** | .586** | .609** | .505** | -- | | | | | | |
| | Access student data | 302 | 4.36 | 2.195 | .385** | .431** | .403** | .340** | .444** | .355** | .443** | .460** | .395** | .409** | .442** | .429** | .507** | -- | | | | | |
| | Regularly update | 302 | 5.16 | 1.903 | .567** | .591** | .564** | .532** | .652** | .570** | .628** | .678** | .601** | .619** | .678** | .458** | .634** | .492** | -- | | | | |
| | Student decision making | 302 | 5.15 | 1.894 | .528** | .544** | .598** | .562** | .673** | .549** | .608** | .615** | .571** | .599** | .705** | .449** | .637** | .464** | .759** | -- | | | |
| | Complete profile | 279 | 5.06 | 1.973 | .512** | .568** | .550** | .575** | .635** | .589** | .642** | .638** | .657** | .610** | .691** | .442** | .579** | .444** | .644** | .626** | -- | | |
| | Skill development | 266 | 503 | 1.932 | .515** | .534** | .575** | .560** | .676** | .555** | .637** | .660** | .587** | .665** | .638** | .452** | .591** | .355** | .564** | .628** | .656** | -- | |
| Ethics and Privacy -Teacher (EPT1- EPT2) | Data protection guidelines | 261 | 6.16 | 1.474 | .372** | .340** | .474** | .426** | .437** | .484** | .396** | .417** | .430** | .404** | .359** | .287** | .462** | .152* * | .332** | .366** | .331** | .392** | -- |
| | Ethical guidelines | 261 | 6.01 | 1.625 | .398** | .359** | .447** | .432** | .493** | .525** | .453** | .511** | .501** | .488** | .412** | .258** | .445** | .265** | .403** | .350** | .381** | .381** | .752** | -- |

** The correlation is significant at the 0.01 level (2-sided). * The correlation is significant at the 0.05 level (2-sided)

Table 5: values and correlation of the students survey for the predicted expectations ('I expect this to happen in reality')

| | | *n* | *M* | *SD* | 1 | 2 | 3 | 4 | 5 | 6 | 7 | 8 | 9 | 10 | 11 | 12 | 13 | 14 | 15 | 16 | 17 | 18 | 19 | 20 |
|---|---|---|---|---|---|---|---|---|---|---|---|---|---|---|---|---|---|---|---|---|---|---|---|---|
| Service Features for Teachers (SFT1-SFT11) | Analytics guidance | 347 | 4.16 | 1.716 | -- | | | | | | | | | | | | | | | | | | | |
| | Access to student progress | 346 | 3.80 | 1.734 | .671** | -- | | | | | | | | | | | | | | | | | | |
| | Professional development | 351 | 4.13 | 1.627 | .769** | .596** | -- | | | | | | | | | | | | | | | | | |
| | Open discussions | 347 | 3.66 | 1.646 | .615** | .568** | .662** | -- | | | | | | | | | | | | | | | | |
| | Improve teaching quality | 261 | 3.62 | 1.688 | .465** | .479** | .458** | .538** | -- | | | | | | | | | | | | | | | |
| | Understandable data and feedback | 279 | 3.60 | 1.764 | .569** | .488** | .519** | .500** | .538** | -- | | | | | | | | | | | | | | |
| | Integrate into feedback | 266 | 3.12 | 1.682 | .480** | .388** | .486** | .496** | .511** | .585** | -- | | | | | | | | | | | | | |
| | Students' learning performance | 266 | 3.55 | 1.837 | .488** | .405** | .406** | .462** | .737** | .565** | .517** | -- | | | | | | | | | | | | |
| | Accurate data | 278 | 3.45 | 1.799 | .441** | .489** | .417** | .489** | .619** | .538** | .433** | .526** | -- | | | | | | | | | | | |
| | Didactic decisions | 261 | 3.51 | 1.755 | .431** | .404** | .433** | .506** | .870** | .529** | .465** | .712** | .547** | -- | | | | | | | | | | |
| | Course goals | 279 | 3.38 | 1.649 | .489** | .484** | .482** | .478** | .589** | .666** | .552** | .553** | .545** | .546** | -- | | | | | | | | | |
| Serv ice Feat ures for | Obligation to act | 266 | 3.07 | 1.748 | .209** | .295** | .313** | .271** | .335** | .262** | .302** | .273** | .286** | .316** | .403** | -- | | | | | | | | |
| | Early interventions | 301 | 3.23 | 1.709 | .450** | .398** | .483** | .499** | .462** | .527** | .533** | .423** | .410** | .423** | .563** | .477** | -- | | | | | | | |
| | Access student data | 302 | 2.88 | 1.823 | .328** | .369** | .372** | .344** | .277** | .418** | .497** | .304** | .329** | .308** | .482** | .456** | .621** | -- | | | | | | |
| | Regularly update | 302 | 3.73 | 1.595 | .504** | .506** | .538** | .497** | .461** | .466** | .424** | .430** | .377** | .419** | .535** | .362** | .517** | .397** | -- | | | | | |
| | Student decision making | 301 | 3.22 | 1.602 | .458** | .485** | .480** | .462** | .599** | .490** | .513** | .499** | .448** | .544** | .549** | .394** | .604** | .516** | .574** | -- | | | | |
| | Complete profile | 279 | 3.66 | 1.814 | .479** | .464** | .485** | .422** | .453** | .519** | .437** | .444** | .533** | .454** | .632** | .499** | .513** | .401** | .497** | .412** | -- | | | |
| | Skill development | 266 | 2.90 | 1.535 | .316** | .307** | .343** | .389** | .520** | .452** | .540** | .498** | .419** | .504** | .467** | .392** | .445** | .401** | .419** | .446** | .452** | -- | | |
| Ethics and Privacy -Teacher (EPT1-EPT2) | Data protection guidelines | 261 | 4.91 | 1.984 | .404** | .332** | .437** | .378** | .553** | .395** | .409** | .409** | .485** | .487** | .345** | .268** | .359** | .192** | .358** | .375** | .428** | .273** | -- | |
| | Ethical guidelines | 261 | 4.43 | 1.944 | .355** | .254** | .420** | .418** | .541** | .436** | .463** | .407** | .431** | .513** | .426** | .339** | .464** | .329** | .412** | .342** | .481** | .303** | .803** | -- |

** The correlation is significant at the 0.01 level (2-sided). * The correlation is significant at the 0.05 level (2-sided)

Table 6: Items of students survey in English and German version with two factors:
Service Features factor (S1-S7) and Ethics and Privacy factor (E1–E5)

| ID Stud-ents | Factor ID | Items student questionnaire | |
|---|---|---|---|
| | | English version | German version |
| 1 | E1 | **Identifiable data**: The university will ask for my consent before using any identifiable data about myself (e.g., ethnicity, age, and gender). | **Identifiable data**: Die Universität wird mich um Erlaubnis bitten, bevor sie Daten von mir nutzt, mit denen ich identifiziert werden kann (z.B. ethnische Zugehörigkeit, Alter und Geschlecht). |
| 2 | E2 | **Keep data secure:** The university will ensure that all my educational data will be kept securely. | **Keep data secure:** Die Universität wird sicherstellen, dass alle meine bildungsbezogenen Daten sicher aufbewahrt werden. |
| 3 | E3 | **Third party**: The university will ask for my consent before my educational data is outsourced for analysis by third party companies. | **Third party**: Die Universität wird um meine Zustimmung bitten, bevor meine bildungsbezogenen Daten zur Analyse an Drittunternehmen ausgelagert werden. |
| 5 | E4 | **Consent to collect:** The university will ask for my consent to collect, use, and analyse any of my educational data (e.g., grades, attendance, and virtual learning environment accesses). | **Consent to collect:** Die Universität wird um meine Zustimmung zur Erhebung, Nutzung und Analyse meiner bildungsbezogenen Daten bitten. |
| 6 | E5 | **Alternative purpose**: The university will request further consent if my educational data is being used for a | **Alternative purpose**: Die Universität wird eine zusätzliche Einwilligung einholen, wenn meine bildungsbezogenen Daten für |
| | | purpose different to what was originally stated. | einen anderen Zweck als den ursprünglich angegebenen verwendet werden. |
| 4 | S1 | **Regularly update**: The university will regularly update me about my learning progress based on the analysis of my educational data. | **Regularly update**: Die Universität wird mich regelmäßig auf Basis meiner bildungsbezogenen Daten über meine Lernfortschritte informieren. |
| 7 | S2 | **Student decision making**: The learning analytics service will be used to promote student decision making (e.g., encouraging you to adjust your set learning goals based upon the feedback provided to you and draw your own conclusions from the outputs received). | **Student decision making**: Der Learning Analytics-Service wird genutzt werden, um meine Entscheidungsfindung als Studierende:r zu fördern, z.B. indem ich ermutigt werde, gesetzte Lernziele basierend auf dem Feedback, das ich erhalten habe, anzupassen und meine eigenen Schlussfolgerungen aus den erhaltenen Ergebnissen zu ziehen. |
| 8 | S3 | **Course goals**: The learning analytics service will show how my learning progress compares to my learning goals/the course objectives. | **Course goals**: Der Learning Analytics-Service wird anzeigen, wie sich mein Lernfortschritt im Vergleich zu meinen Lernzielen / den Kurszielen darstellt. |
| 9 | S4 | **Complete profile**: The learning analytics service will present me with a complete profile of my learning across every module (e.g., number of accesses to online material and attendance). | **Complete profile**: Der Learning Analytics-Service wird mir ein vollständiges Profil meines Lernens über jedes Modul hinweg bieten (z.B. Angaben zur Anzahl der Zugriffe auf Online-Material, zu meinen Lernergebnissen und meiner Anwesenheit). |
| 10 | S5 | **Integrate into feedback**: The teaching staff will be competent in incorporating analytics into the feedback and support they provide to me. | **Integrate into feedback**: Die Lehrkräfte werden kompetent sein, die Learning Analytics-Ergebnisse in das Feedback und die Unterstützung, die sie mir geben, einzubeziehen. |
| 11 | S6 | **Obligation to act:** The teaching staff will have an obligation to act (i.e., support me) if the analytics show that I am at-risk of failing, underperforming, or if I could improve my learning. | **Obligation to act:** Die Lehrkräfte werden verpflichtet sein zu handeln, d.h. mich zu unterstützen, wenn die Learning Analytics-Ergebnisse zeigen, dass ich durchzufallen drohe, die Erwartungen nicht zu erfüllen, oder wenn ich mein Lernen verbessern könnte. |
| 12 | S7 | **Skill development:** The feedback from the learning analytics service will be used to promote academic and professional skill development (e.g., essay writing and referencing) for my future employability. | **Skill development:** Die Ergebnisse des Learning-Analytics-Services werden genutzt werden, um die akademische und berufliche Kompetenzentwicklung (z.B. Verfassen von Texten und Zitieren von Literatur) für meine zukünftige Beschäftigungsfähigkeit zu fördern. |

Table 7: Items of teacher survey in English and German version with three factors:
Service Features for Teachers (SFT1-SFT11), Service Features for Students (SFS1-SFS7)

and Ethics and Privacy – Teacher (EPT1-EPT2)

| ID Teacher | ID sub-scale Teacher | Items teacher questionnaire | |
|---|---|---|---|
| | | English version | German version |
| **2** | SFT1 | **Analytics guidance**: The university will provide me with guidance on how to access learning analytics about my students. | **Analytics guidance:** Die Universität wird mir Anleitungen zur Verfügung stellen, wie ich auf die Learning Analytics-Ergebnisse über meine Studierenden zugreifen kann. |
| **3** | SFT2 | **Access to student progress:** I will be able to access data about my students' progress in a course that I am teaching/tutoring. | **Access to student progress:** Ich werde die Möglichkeit haben, auf Lernprozessdaten der Studierenden in den Lehrveranstaltungen, in denen ich lehre, zuzugreifen. |
| **1** | SFT3 | **Professional development:** The University will provide staff with opportunities for professional development in using learning analytics for teaching. | **Professional development:** Die Universität wird uns Lehrenden Weiterbildungsangebote zur Verfügung stellen, wie Learning Analytics in der Lehre eingesetzt werden kann. |
| **4** | SFT4 | **Open discussions:** The university will facilitate open discussions to share experience of learning analytics services. | **Open discussions:** Die Universität wird den offenen Erfahrungsaustausch zwischen allen Beteiligten im Kontext von Learning Analytics fördern. |
| **19** | SFT5 | **Improve teaching quality:** The results from the Learning Analytics will help me to improve the teaching quality of my course (e.g. identification of problems in feedback, exams and learning activities). | **Improve teaching quality:** Die Ergebnisse aus den Learning Analytics werden mir helfen, die Qualität meiner Lehre im Kurs zu verbessern (z. B. Identifikation von Problemen beim Feedback, bei Prüfungen und zu Lernaktivitäten). |
| **9** | SFT6 | **Understandable data and feedback:** The feedback from the learning analytics service will be presented in a format that is both understandable and easy to read. | **Understandable data and feedback:** Die Ergebnisse aus dem Learning Analytics-Service werden in einer leicht verständlichen und einfach zu lesenden Art dargestellt. |
| **13** | SFT7 | **Integrate into feedback**: The teaching staff will be competent in incorporating analytics into the feedback and support they provide to students. | **Integrate into feedback**: Das Lehrpersonal wird kompetent darin sein, die Learning Analytics-Ergebnisse mit in das Feedback und in die Unterstützung ihrer Studierenden einzubeziehen. |
| **16** | SFT8 | **Students' learning performance**: The use of learning analytics will allow me to better understand my students' learning performance. | **Students' learning performance:** Die Nutzung von Learning Analytics wird mir helfen, die Lernleistung meiner Studierenden besser zu verstehen. |
| **12** | SFT9 | **Accurate data:** The learning analytics service will collect and present data that is accurate (i.e., free from inaccuracies such as incorrect grades). | **Accurate data:** Der Learning Analytics-Service wird Daten sammeln und präsentieren, die korrekt sind (d.h. frei von Ungenauigkeiten wie z.B. fehlerhafte Noten). |
| **20** | SFT10 | **Didactic decisions:** The Learning Analytics service will enable me to make didactic decisions (e.g., method selection) based on the Learning Analytics results. | **Didactic decisions:** Der Learning Analytics-Service wird es mir ermöglichen, auf Grundlage der Learning Analytics-Ergebnisse didaktische Entscheidungen (z. B. Methodenauswahl) zu treffen. |
| **10** | SFT11 | **Course goals**: The learning analytics service will show how a student's learning progress compares to their learning goals/the course objectives. | **Course goals**: Der Learning Analytics-Service wird anzeigen, wie sich der Lernfortschritt der Studierenden im Vergleich zu meinen Lehrzielen / den Kurszielen darstellt. |
| **15** | SFS1 | **Obligation to act:** The teaching staff will have an obligation to act (i.e., support students) if the analytics show that a student is at-risk of failing, underperforming, or that they could improve their learning. | **Obligation to act:** Das Lehrpersonal wird verpflichtet sein zu handeln (d. h. die Studierenden zu unterstützen), wenn die Learning Analytics-Ergebnisse zeigen, dass ein Student oder eine Studentin durchzufallen droht, die Erwartungen nicht erfüllt oder ihre / seine Lernleistung verbessern könnte. |
| **7** | SFS2 | **Early interventions:** The university will provide support (e.g., advice from personal tutors) as soon as possible if the analysis of a student's educational data suggests they may be having some difficulty or problem (e.g., underperforming or at-risk of failing). | **Early interventions:** Wenn die Analyse von Lerndaten darauf hinweist, dass Studierende Probleme haben, z. B. da sie hinter den Erwartungen zurückbleiben oder drohen durchzufallen, wird die Universität schnellstmöglich Unterstützungsmaßnahmen anbieten (z.B. Studienberatungsangebote). |
| **8** | SFS3 | **Access student data:** I will be able to access data about any students within a programme. | **Access student data:** Ich werde die Möglichkeit haben, auf Lernprozessdaten aller Studierenden in einem Studiengang zuzugreifen. |
| **5** | SFS4 | **Regularly update:** The university will regularly update students about their learning progress based on the analysis of their educational data. | **Regularly update:** Die Universität wird die Studierenden regelmäßig auf Basis ihrer bildungsbezogenen Daten über ihren Lernfortschritt informieren. |
| **6** | SFS5 | **Student decision making:** The learning analytics service will allow students to make their own decisions based on the data they receive. | **Student decision making**: Der Learning Analytics-Service wird es den Studierenden ermöglichen, auf Grundlage ihrer Lernprozessdarstellung eigene Entscheidungen zu treffen. |
| **11** | SFS6 | **Complete profile**: The learning analytics service will present students with a complete profile of their | **Complete profile**: Der Learning Analytics-Service wird den Studierenden ein vollständiges Profil |
| | | learning across every course (e.g., number of accesses to online material, learning outcomes, and attendance). | ihres Lernens über alle Kurse hinweg anzeigen (z. B. Angaben zu Anzahl der Zugriffe auf Online-Material, zu ihren Lernergebnissen und ihrer Anwesenheit). |
| **14** | SFS7 | **Skill development:** The feedback from the learning analytics service will be used to promote students' academic and professional skill development (e.g., essay writing and referencing) for their future employability. | **Skill development:** Die Ergebnisse aus dem Learning Analytics-Service werden auch dazu genutzt, um die akademische und berufliche Kompetenzentwicklung der Studierenden (z.B. Verfassen von Texten und Zitieren von Literatur) für ihre zukünftige Beschäftigungsfähigkeit zu fördern. |
| **18** | EPT1 | **Data protection guidelines:** The university will provide clear data protection guidelines for the use of learning analytics. | **Data protection guidelines:** Die Universität wird eindeutige datenschutzrechtliche Vorgaben für die Nutzung von Learning Analytics machen. |
| **17** | EPT2 | **Ethical guidelines:** The university will set clear ethical guidelines for the use of learning analytics. | **Ethical guidelines:** Die Universität wird eindeutige ethische Vorgaben für die Nutzung von Learning Analytics machen. |

Table 8: Comparable items of the student and teacher survey in English and German version

| ID Stud-ents | ID sub-scale Stud-ents | ID Tea-cher | ID sub-scale Teacher | Items student questionnaire | | Items teacher questionnaire | |
|---|---|---|---|---|---|---|---|
| | | | | English version | German version | English version | German version |
| 4 | S1 | 5 | SFS4 | **Regularly update**: The university will regularly update me about my learning progress based on the analysis of my educational data. | **Regularly update**: Die Universität wird mich regelmäßig auf Basis meiner bildungsbezogenen Daten über meine Lernfortschritte informieren. | **Regularly update:** The university will regularly update students about their learning progress based on the analysis of their educational data. | **Regularly update:** Die Universität wird die Studierenden regelmäßig auf Basis ihrer bildungsbezogenen Daten über ihren Lernfortschritt informieren. |
| 7 | S2 | 6 | SFS5 | **Student decision making**: The learning analytics service will be used to promote student decision making (e.g., encouraging you to adjust your set learning goals based upon the feedback provided to you and draw your own conclusions from the outputs received). | **Student decision making**: Der Learning Analytics-Service wird genutzt werden, um meine Entscheidungsfindung als Studierende:r zu fördern, z.B. indem ich ermutigt werde, gesetzte Lernziele basierend auf dem Feedback, das ich erhalten habe, anzupassen und meine eigenen Schlussfolgerungen aus den erhaltenen Ergebnissen zu ziehen. | **Student decision making:** The learning analytics service will allow students to make their own decisions based on the data they receive. | **Student decision making**: Der Learning Analytics-Service wird es den Studierenden ermöglichen, auf Grundlage ihrer Lernprozessdarstellung eigene Entscheidungen zu treffen. |
| 8 | S3 | 10 | SFT11 | **Course goals**: The learning analytics service will show how my learning progress compares to my learning goals/the course objectives. | **Course goals**: Der Learning Analytics-Service wird anzeigen, wie sich mein Lernfortschritt im Vergleich zu meinen Lernzielen / den Kurszielen darstellt. | **Course goals**: The learning analytics service will show how a student's learning progress compares to their learning goals/the course objectives. | **Course goals**: Der Learning Analytics-Service wird anzeigen, wie sich der Lernfortschritt der Studierenden im |
| | | | | | | | Vergleich zu meinen Lehrzielen / den Kurszielen darstellt. |
| 9 | S4 | 11 | SFS6 | **Complete profile**: The learning analytics service will present me with a complete profile of my learning across every module (e.g., number of accesses to online material and attendance). | **Complete profile**: Der Learning Analytics-Service wird mir ein vollständiges Profil meines Lernens über jedes Modul hinweg bieten (z.B. Angaben zur Anzahl der Zugriffe auf Online-Material, zu meinen Lernergebnissen und meiner Anwesenheit). | **Complete profile**: The learning analytics service will present students with a complete profile of their learning across every course (e.g., number of accesses to online material, learning outcomes, and attendance). | **Complete profile**: Der Learning Analytics-Service wird den Studierenden ein vollständiges Profil ihres Lernens über alle Kurse hinweg anzeigen (z. B. Angaben zu Anzahl der Zugriffe auf Online-Material, zu ihren Lernergebnissen und ihrer Anwesenheit). |
| 10 | S5 | 13 | SFT7 | **Integrate into feedback**: The teaching staff will be competent in incorporating analytics into the feedback and support they provide to me. | **Integrate into feedback**: Die Lehrkräfte werden kompetent sein, die Learning Analytics-Ergebnisse in das Feedback und die Unterstützung, die sie mir geben, einzubeziehen. | **Integrate into feedback**: The teaching staff will be competent in incorporating analytics into the feedback and support they provide to students. | **Integrate into feedback**: Das Lehrpersonal wird kompetent darin sein, die Learning Analytics-Ergebnisse mit in das Feedback und in die Unterstützung ihrer Studierenden einzubeziehen. |
| 11 | S6 | 15 | SFS1 | **Obligation to act:** The teaching staff will have an obligation to act (i.e., support me) if the analytics show that I am at-risk of failing , underperforming, or if I could improve my learning. | **Obligation to act:** Die Lehrkräfte werden verpflichtet sein zu handeln, d.h. mich zu unterstützen, wenn die Learning Analytics-Ergebnisse zeigen, dass ich durchzufallen drohe, die Erwartungen nicht zu erfüllen, oder wenn ich mein Lernen verbessern könnte. | **Obligation to act:** The teaching staff will have an obligation to act (i.e., support students) if the analytics show that a student is at-risk of failing, underperforming, or that they could improve their learning. | **Obligation to act:** Das Lehrpersonal wird verpflichtet sein zu handeln (d. h. die Studierenden zu unterstützen), wenn die Learning Analytics-Ergebnisse zeigen, dass ein Student oder eine Studentin durchzufallen droht, die Erwartungen nicht erfüllt oder ihre / seine Lernleistung verbessern könnte. |
| 12 | S7 | 14 | SFS7 | **Skill development:** The feedback from the learning analytics service will be used to promote academic and professional skill development (e.g., essay writing and referencing) for my future employability. | **Skill development:** Die Ergebnisse des Learning-Analytics-Services werden genutzt werden, um die akademische und berufliche Kompetenzentwicklung (z.B. Verfassen von Texten und Zitieren von Literatur) für meine zukünftige Beschäftigungsfähigkeit zu fördern. | **Skill development:** The feedback from the learning analytics service will be used to promote students' academic and professional skill development (e.g., essay writing and referencing) for their future employability. | **Skill development:** Die Ergebnisse aus dem Learning Analytics-Service werden auch dazu genutzt, um die akademische und berufliche Kompetenzentwicklung der Studierenden (z.B. Verfassen von Texten und Zitieren von Literatur) für ihre zukünftige Beschäftigungsfähigkeit zu fördern. |

Table 9: Factor analysis of student sample

| Factor 1: Ethics and Privacy factor (E1–E5) | Explained variance in% |
|---|---|
| | 24.831 |
| | factor score |
| Identifiable data | .705 |
| Keep data secure | .723 |
| Third party | .786 |
| Consent to collect | .822 |
| Alternative purpose | .796 |
| Factor 2: Service Features factor (S1-S7) | Explained variance in% |

| | 36.035<br>factor score |
|---|---|
| Regularly update | .707 |
| Student decision making | .845 |
| Course goals | .835 |
| Complete profile | .817 |
| Integrate into feedback | .754 |
| Obligation to act | .732 |
| Skill development | .795 |

Table 10: Factor analysis of the teacher sample

| Factor 1: Service Features for Teachers (SFT1-SFT11) | Explained variance in% | Factor 2: Service Features for students (SFS1-SFS7) | Explained variance in% |
|---|---|---|---|
| | 32.77<br>factor score | | 23.90<br>factor score |
| Analytics guidance | .859 | Obligation to act | .756 |
| Access to student progress | .803 | Early interventions | .739 |
| Professional development | .764 | Access student data | .656 |
| Open discussions | .743 | Regularly update | .635 |
| Improve teaching quality | .667 | Student decision making | .624 |
| Understandable data and feedback | .646 | Complete profile | .582 |
| Integrate into feedback | .637 | Skill development | .545 |
| Students' learning performance | .613 | Factor 3: Ethics and Privacy -Teacher (EPT1-EPT2) | Explained variance in% |
| Accurate data | .597 | | 12.79 |
| Didactic decisions | .592 | | factor score |
| Course goals | .571 | Data protection guidelines | .885 |
| | | Ethical guidelines | .855 |

Table 11: Mean values and mean difference of the student sample

| | | *M* ideal | *M* predict | *M* Diff. |
|---|---|---|---|---|
| Ethics and privacy (E1 – E5) | Identifiable data | 6.35 | 5.15 | 1.2 |
| | Keep data secure | 6.65 | 5.12 | 1.53 |
| | Third party | 6.60 | 4.89 | 1.71 |
| | Consent to collect | 6.33 | 4.94 | 1.39 |
| | Alternative purpose | 6.59 | 4.75 | 1.84 |
| Service features (S1–S7) | Regularly update | 5.12 | 3.4 | 1.72 |
| | Student decision making | 5.48 | 3.8 | 1.68 |
| | Course goals | 5.47 | 3.91 | 1.56 |
| | Complete profile | 5.08 | 3.89 | 1.19 |
| | Integrate into feedback | 5.74 | 3.17 | 2.57 |
| | Obligation to act | 4.95 | 2.91 | 2.04 |
| | Skill development | 5.30 | 3.47 | 1.83 |

Table 12: Mean values and mean difference of the teacher sample

| | | *M* ideal | *M* predict | *M* Diff. |
|---|---|---|---|---|
| Service Features for Teachers (SFT1-SFT11) | Analytics guidance | 5.59 | 4.16 | 1.43 |
| | Access to student progress | 5.30 | 3.80 | 1.50 |
| | Professional development | 5.53 | 4.13 | 1.40 |
| | Open discussions | 5.36 | 3,66 | 1.70 |
| | Improve teaching quality | 5.56 | 3.62 | 1.94 |
| | Understandable data and feedback | 5.91 | 3.60 | 2.31 |
| | Integrate into feedback | 5.63 | 3.12 | 2.51 |
| | Students' learning performance | 5.45 | 3.55 | 1.90 |
| | Accurate data | 5.68 | 3.45 | 2.23 |
| | Didactic decisions | 5.24 | 3.51 | 1.73 |
| | Course goals | 5.30 | 3.38 | 1.92 |
| Service Features for students (SFS1- FS7) | Obligation to act | 3.74 | 3.07 | 0.67 |
| | Early interventions | 5.29 | 3.23 | 2.06 |
| | Access student data | 4.36 | 2.88 | 1.48 |
| | Regularly update | 5.16 | 3.73 | 1.43 |
| | Student decision making | 5.15 | 3.22 | 1.93 |
| | Complete profile | 5.06 | 3.66 | 1.40 |
| | Skill development | 5.03 | 2.90 | 2.13 |
| Ethics and Privacy -Teacher (EPT1-EPT2) | Data protection guidelines | 6.16 | 4.91 | 1.25 |
| | Ethical guidelines | 6.01 | 4.43 | 1.58 |

Table 13: Mean values and mean differences of the comparable items

| | *M* student 'ideal' | *M* students 'predicted' | *M* difference students | *M* teachers 'ideal' | *M* teachers 'predicted' | *M* difference teachers |
|---|---|---|---|---|---|---|
| Regularly update | 5.12 | 3.40 | 1.72 | 5.16 | 3.73 | 1.43 |
| Student decision making | 5.48 | 3.80 | 1.68 | 5.15 | 3.22 | 1.93 |
| Course goals | 5.47 | 3.91 | 1.56 | 5.30 | 3.38 | 1.92 |
| Complete profile | 5.08 | 3.89 | 1.19 | 5.06 | 3.66 | 1.4 |
| Integrate into feedback | 5.74 | 3.17 | 2.57 | 5.63 | 3.12 | 2.51 |
| Obligation to act | 5.30 | 3.47 | 1.83 | 5.03 | 2.90 | 2.13 |
| Skill development | 4.95 | 2.91 | 2.04 | 3.74 | 3.07 | 0.67 |